\documentclass{aa}  
\usepackage{graphicx}
\usepackage{txfonts}
\usepackage{placeins}
\usepackage{xcolor} 
\newcommand{\asec}{$^{\prime\prime}$}

\def\NH2{$N\rm{(H_2)}$}

\def\SigmaH2{$\Sigma $(${\rm H_2}$)}
\def\r1415{$^{14}$N/$^{15}$N}

\def\cyclic{{\it c-}C$_3$H$_2$}

\def\15N{$^{15}$NNH$^+$}
\def\N15{N$^{15}$NH$^+$}

\def\HCOp{\mbox{HCO$^+$}}
\def\HCOpI{\mbox{H$^{13}$CO$^+$}}

\def\METH{CH$_3$OH}
\def\FORM{H$_2$CO}

\def\kms{\mbox{km~s$^{-1}$}}
\def\cmc{cm$^{-3}$}
\def\cmq{cm$^{-2}$}

\def\solm{\mbox{M$_\odot$}}

\def\Tex{\mbox{$T_{\rm ex}$}}

\def\Ntot{\mbox{$N_{\rm tot}$}}
\def\thetaS{\mbox{$\theta_{\rm s}$}}

\def\Tb{\mbox{$T_{\rm B}$}}

\def\Tex{\mbox{$T_{\rm ex}$}}

\def\kms{km\,s$^{-1}$}

\begin{document}

   \title{CHEMOUT: CHEMical complexity in star-forming regions of the OUTer Galaxy}

   \subtitle{IV. ALMA observations of organic species at Galactocentric radius $\sim 23$~kpc}

   \author{F. Fontani
          \inst{1,2,3}
          \and
          G. Vermari\"{e}n\inst{4}
          \and
          S. Viti\inst{4}
          \and
          D. Gigli\inst{1,5}
          \and
          L. Colzi\inst{6}
          \and
          M.T. Beltr\'an\inst{1}
          \and
          P. Caselli\inst{2}
          \and
          V.M. Rivilla\inst{6}
          \and
          A. S\'anchez-Monge\inst{7,8}
          }
    \institute{INAF - Osservatorio Astrofisico di Arcetri,
              Largo E. Fermi 5,
              I-50125, Florence, Italy \\
              \email{francesco.fontani@inaf.it}
              \and
              Max-Planck-Institut f\"{u}r extraterrestrische Physik, Giessenbachstra{\ss}e 1, 85748 Garching bei M\"{u}nchen, Germany
              \and
              LERMA, Observatoire de Paris, PSL Research University, CNRS, Sorbonne
              Universit\'e, F-92190 Meudon, France
              \and
              Leiden Observatory, Leiden University, PO Box 9513, 2300 RA Leiden, The Netherlands
              \and
             Leiden Observatory, Leiden University, PO Box 9513, 2300 RA Leiden, The Netherlands
             \and 
              Dipartimento di Fisica e Astronomia, Universit\`a di Firenze, Via G. Sansone 1, 50019 Sesto Fiorentino, Firenze, Italy
              \and
              Centro de Astrobiolog\'ia (CSIC-INTA), Ctra Ajalvir km 4, 28850, Torrej\'on de Ardoz, Madrid, Spain
             \and
      Institut de Ci\`encies de l'Espai (ICE, CSIC), Campus UAB, Carrer de Can Magrans s/n, 08193, Bellaterra (Barcelona), Spain
      \and
      Institut d'Estudis Espacials de Catalunya (IEEC), 08860 Castelldefels (Barcelona), Spain
              }
   \date{Received XXX; accepted XXX}

 
  \abstract
   {Single-dish observations suggest that the abundances of organic species in star-forming regions of the outer Galaxy, characterised by sub-Solar metallicities, are comparable to those found in the local Galaxy.}
   {To understand this counter-intuitive result, and avoid misleading interpretation due to beam dilution effects at such large distances, spatially resolved molecular emission maps are needed to link correctly measured abundances and local physical properties.}
   {We observed several organic molecules with the Atacama Large Millimeter Array towards WB89-671, the source with the largest Galactocentric distance (23.4~kpc) of the project "CHEMical complexity in star-forming regions of the OUTer Galaxy" (CHEMOUT), at a resolution of $\sim 15000$~au. 
   We compared the observed molecular abundances with chemical model predictions.}
   {We detected emission of \cyclic, C$_4$H, \METH, \FORM, HCO, \HCOpI, HCS$^+$, CS, HN$^{13}$C, and SO.
   The emission morphology is complex, extended, and different in each tracer.
   In particular, the most intense emission in \HCOpI, \FORM\ and \cyclic\ arises from two millimeter continuum, infrared-bright cores.
   The most intense \METH\ and SO emission arises predominantly from the part of the filament with no continuum sources.
   The narrow linewidths across the filament indicate quiescent gas, despite the two embedded protostars.
   Derived molecular column densities are comparable with those in local star-forming regions, and suggest anti-correlation between hydrocarbons, ions, HCO, and \FORM\ on one side, and \METH\ and SO on the other.}
  {Static chemical models that best match the observed column densities favour low energetic conditions, expected at large Galactocentric radii, but carbon elemental abundances 3 times higher than that derived extrapolating the [C/H] Galactocentric gradient at 23~kpc. 
  This would indicate a flatter [C/H] trend at large Galactocentric radii, in line with a flat abundance of organics.
   However, to properly reproduce the chemical composition of each region, models should include dynamical evolution.}


   \maketitle
%

\section{Introduction}
\label{intro}

The outer Galaxy (OG), i.e. the part of the Milky Way that extends out of the Solar circle to the outermost edge of the Galactic disc ($\sim 27$~kpc,
L\'opez-Corredoira et al. 2018), was believed to be an environment not optimal for the formation of 
molecules and planetesimals. The reason is its sub-Solar metallicity, i.e. the low abundance of elements heavier 
than helium. 
In particular, the elemental abundances of oxygen, carbon, and nitrogen, i.e. the three 
most abundant elements in the Universe after hydrogen and helium, and the most important biogenic elements, 
decrease as a function of the Galactocentric distance, $R_{\rm GC}$ \citep{esteban17,arellano20,mendez22}. 
For example, the fractional abundance of oxygen relative to hydrogen, [O/H], decreases gradually from the inner Galaxy
to the OG, reaching $\sim$1/5th of the Solar value at about 20 kpc \citep[see e.g.][]{esteban17,mendez22}. 
For carbon, the decrease of the [C/H] ratio is even more pronounced, reaching $\sim$1/7th--1/8th of the Solar value at $\sim$20 kpc \citep{arellano20,mendez22}.
Such low abundances of heavy elements in the OG suggested in the past that this zone was not suitable for forming planetary 
systems in which Earth-like planets could be born and might be capable of sustaining life \citep{ramirez10}.
For this reason, the OG was excluded from the so-called Galactic Habitable Zone (GHZ), 
which is the portion of the Milky Way with the highest chance to form and develop complex forms of life 
on (exo-)planets \citep{gonzalez01}. Because of this, its chemical complexity has been so far little explored. 

However, we are discovering that the presence of small, terrestrial 
planets is independent on the Galactocentric distance \citep[e.g.][]{buchhave12,meb20}, 
and that molecules, including complex organic molecules (COMs, organic species with 6 or more atoms), are found to be more-than-expected abundant in star-forming 
regions with metallicity lower than Solar, both in the OG \citep[e.g.][]{blair08,shimonishi21, bernal21,fontani22a} and in external galaxies \citep[e.g.][]{shimonishi18,sewilo18}.
However, molecular formation processes could be different from those in the inner or local Galaxy due to the difference
in both metallicity and other environmental conditions. 
For example, UV irradiation from high-mass stars should be lower, on average, in the OG due to the smaller concentration of high-mass stars.
Therefore, abundances of species sensitive to this parameter could be affected also by this environmental change.
It is thus crucial to observe molecules in star-forming regions of the OG to constrain models 
adapted to such sub-Solar metallicity environments.

The few studies performed so far in the OG mentioned above have provided abundances only of a limited number of abundant species, and hence they can answer only partial questions.
\citet{blair08}, \citet{bernal21}, and \citet{fontani22b} detected formaldehyde (H$_2$CO) and methanol (CH$_3$OH) in star-forming regions of the OG up to 24~kpc. 
H$_2$CO is an important precursor of CH$_3$OH, the simplest complex organics,
because in cold star-forming cores its formation proceeds on the surface of dust grains via successive hydrogenation of CO 
\citep[HCO $\rightarrow$ \FORM\ $\rightarrow$ CH$_3$O/CH$_2$OH $\rightarrow$ \METH, e.g.][]{peg18}. 
However, \FORM\ can also form in 
the gas-phase, unlike \METH, in regions where a significant fraction of C is not yet locked into CO \citep{ramal21}. 
If so, \FORM\ should form together with species like carbon-chains, which need a large amount of C not locked in CO as well. 
This implies that in regions where the C/O ratio is smaller, most of the C should be in the 
form of CO, and carbon-chains should have a lower abundance than that of \METH. 
The OG is an environment that should have this property,
because the [C/H] decrease with $R_{\rm GC}$ is steeper than that of [O/H] \citep{arellano20,mendez22}.
The vice-versa is expected in the inner Galaxy. 
On the other hand, a dust extinction lower in the OG than in the local Galaxy would imply an easier dissociation of CO in gas phase, favouring the formation of carbon chains. 
From this example, one can see that 
the formation of even a relatively simple organic molecule like \FORM, and its relation with \METH, can be different 
in the outer and inner Galaxy. 

The project "CHEMical complexity of star-forming regions in the OUTer Galaxy \citep[CHEMOUT][]{fontani22a,fontani22b,colzi22} aims at studying the formation of molecules in the outer Galaxy based on observations of 35 molecular cloud cores associated with star-forming regions 
having $R_{\rm GC}$ in between ~9 and ~23 kpc.
Using the Institut de Radioastronomie Millim\'etrique (IRAM) 30m telescope, we detected several simple and complex carbon-bearing 
molecules including organics (\cyclic, \HCOp, \HCOpI, HCO, C$_4$H, HCS$^+$, HCN, CH$_3$CCH).
Inorganic tracers of star-formation activity (SO, SiO, N$_2$D$^+$) were detected as well.
The detection of all these species should better constrain their formation/destruction pathways, and 
highlight similarities and differences with those known to be efficient in the local/inner 
Galaxy, where the C/O ratio is different. 
However, the results of \citet{blair08}, \citet{bernal21}, and \citet{fontani22a,fontani22b} are based on observations of 
single-dish telescopes, and thus provide only abundances averaged over angular scales of their main beams ($\sim 27-63$\asec). 
At the distance of the CHEMOUT targets, i.e. 8--15~kpc from the Sun \citep{fontani22a}, an angular scale of 27\asec\ corresponds to 
$1-2$~pc, or 200\,000 -- 400\,000 au, that is at least a factor 10 larger than the typical linear scale of a single star-forming core (0.05--0.1~pc). 

\begin{figure}[h!]
    \centering
    \includegraphics[width=1\hsize]{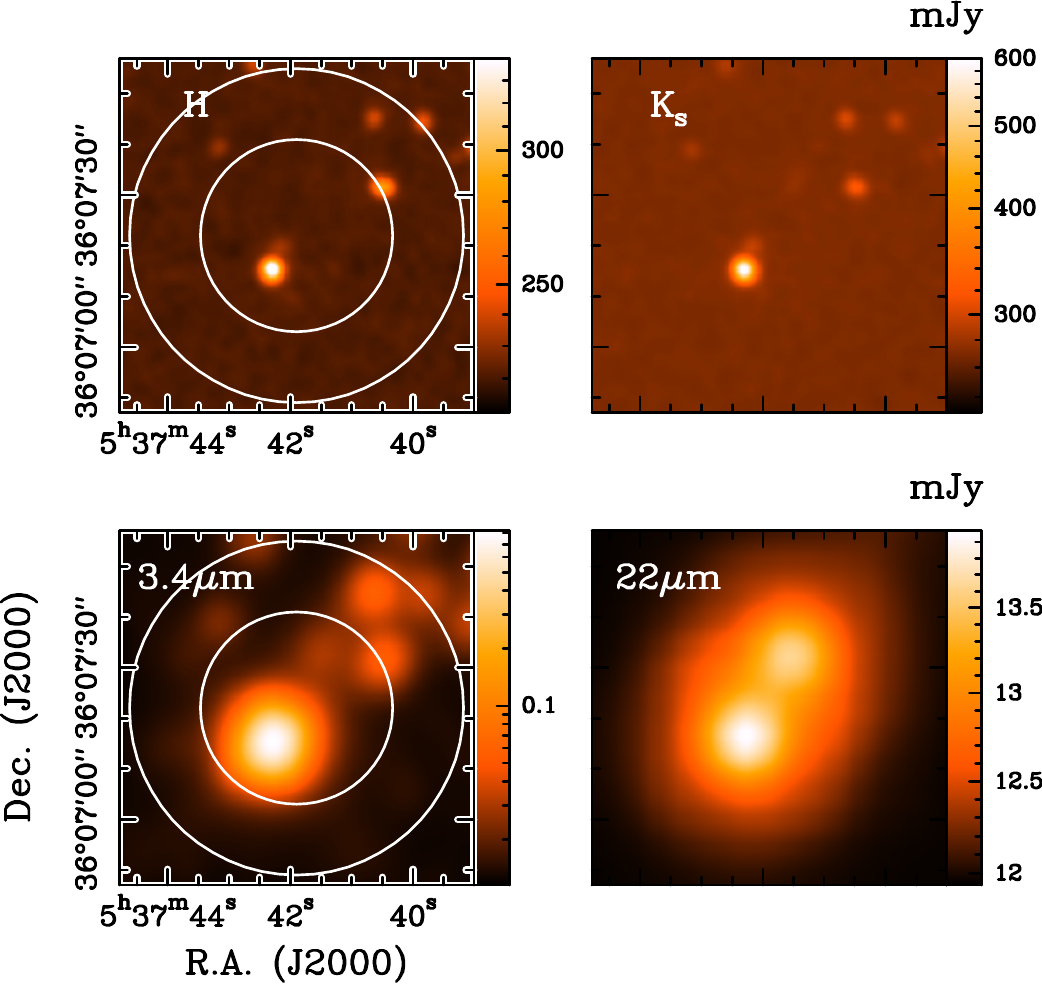}
    \caption{Near- and mid-infrared images of WB89-670. The top row panels show the images at 1.25~$\mu$m (H) and 2.15~$\mu$m (K$_s$) from the 2MASS survey. The panels in the bottom row show the WISE images at the  wavelengths indicated in the top-left corner. The white circles are the ALMA primary beams at $\sim 86$~GHz and $\sim 150$~GHz, that are 66\asec\ (i.e. $\sim 7.5$~pc) and 38\asec\ (i.e. $\sim 4.3$~pc), respectively.}
    \label{fig:2mass}
\end{figure}

In this work, we present high-angular resolution images obtained with the Atacama Large Millimeter Array (ALMA) to resolve the molecular emission towards the source WB89-670 \citep[IRAS 05343+3605][]{web89}, WB670 hereafter. 
The source is located at a $R_{\rm GC}$ of $\sim 23.4$~kpc \citep[heliocentric distance $\sim 15.1$~kpc,][]{fontani22a}, in direction of the Galactic anticentre (Galactic coordinates Lon.=173.014$^{\circ}$, Lat.=2.38$^{\circ}$).  
This is the CHEMOUT target with the largest Galactocentric distance, thus associated with the smallest environmental [C/H] and [O/H] of the sample, as well as with the lowest C/O ratio \citep[][]{esteban17,arellano20,mendez22}.
Extrapolating the elemental Galactocentric gradients of carbon and oxygen measured by \citet{mendez22} to 23.4~kpc, the fractional abundances [C/H] and [O/H] are 1.8$\times$10$^{-5}$ and 6.7$\times$10$^{-5}$, respectively.
Considering the Solar values of [C/H]$\sim 2.6 \times 10^{-4}$ and [O/H]$\sim 3.1\times 10^{-4}$, the [C/O] ratio should be $\sim 3.1$ times lower than Solar in the natal cloud of WB670.
The source harbours near- and mid-infrared sources detected in the images of the Two Micron All-Sky Survey \citep[2MASS,][]{cutri03} and the Wide-field Infrared
Survey Explorer \citep[WISE,][Fig.~\ref{fig:2mass}]{wright10}, associated with molecular gas where rotational transitions of \cyclic, C$_4$H, CCS, \HCOpI, \HCOp, and HCN \citep{fontani22a}, \METH\ \citep{bernal21}, and \FORM\ \citep{blair08} were detected. 
No COMs except \METH\ were identified in this source.
The narrow width at half maximum (about 1~\kms) in the line profile of \HCOp $J=1-0$ suggests 
that the bulk emission in this transition is from quiescent material, but the simultaneous detection of non-Gaussian wings at high velocities also indicates the presence of embedded protostellar activity \citep{fontani22a}.

The paper is organised as follows: the observations and the data reduction are described in Sect.~\ref{obs}. The observational results are shown in Sect.~\ref{res}.
The spectral analysis and derivation of the molecular column densities is illustrated in Sect.~\ref{coldens}.
A discussion of the results and a comparison with chemical mdelling is provided in Sect.~\ref{discu}. Conclusions and future perspectives are given in Sect.~\ref{conc}.

\FloatBarrier

\section{Observations and data reduction}
\label{obs}

\begin{table}[h!]
\begin{center}
\caption{\label{tab:setup} Spectral setup and observational parameters.}
\begin{tabular}{c c c c c c}
\hline
\hline
  spw & $\nu_c$\tablefootmark{a} & $\Delta \nu$\tablefootmark{b} & $\Delta V$\tablefootmark{b} & Chan. & rms\tablefootmark{c} \\
 &  [GHz]  &   [kHz]      &   [\kms]   &     & [mJy beam$^{-1}$] \\
\hline
\multicolumn{6}{c}{Band 3}   \\
 25  &  84.72719 & 122  &  0.43 & 480   &  3.9 \\
 27  &  84.52064 & 122  &  0.43 & 480   &  3.9 \\
 29  &  85.65553 & 122  &  0.43 & 480   &  3.8 \\
 31  &  85.33839 & 122  &  0.43 & 480   &  3.9 \\
 33  &  86.64949 & 488  &  1.72  & 1920  &  1.8 \\
 35  &  97.59944 & 122  &  0.38 & 960   &  3.6 \\
 37  &  96.79341 & 122  &  0.38 & 960   &  3.7 \\
 39  &  97.92939 & 488  &  1.52 & 960   &  1.9 \\
 41  &  99.38944 & 488  &  1.52 & 960   &  2.0 \\
\hline
     \multicolumn{6}{c}{Band 4}   \\
  25  &  139.75200 & 488  & 1.05 &  1920  &  8.8 \\
  27  &  141.83430 & 122  & 0.26 &  480   &  17 \\
  29  &  140.30817 & 122  & 0.26 &  480   &  17 \\
  31  &  140.80201 & 122  & 0.26 &  960   &  16 \\
  33  &  151.38080 & 122  & 0.24 &  480   &  17 \\
  35  &  150.99756 & 122  & 0.24 &  480   &  18 \\
  37  &  150.50049 & 122  & 0.24 &  960   &  18 \\
  39  &  153.45220 & 488  & 0.96 &  1920   &  10 \\
\hline
\end{tabular}
\end{center}
\tablefoot{
\tablefoottext{a}{Central frequency of the spectral window (spw);}
\tablefoottext{b}{Spectral resolution in frequency and velocity;}
\tablefoottext{c}{Root mean square noise per channel;}
}
\end{table}

The observations were carried out with ALMA during Cycle 9 in three dates: 
December 27, 2022, using 45 antennas, and January 8 and 9, 2023 (project
2022.1.00911.S, P.I.: F. Fontani), using 41 antennas. 
The phase centre was set to the equatorial coordinates R.A.(J2000)=05$h$37$m$41.9$s$ and Dec(J2000)=36$^{\circ}$07$^{\prime}$22$^{\prime\prime}$. 
The Local Standard of Rest velocity is --17.4~\kms\ \citep{fontani22a}.
We observed several spectral windows in 
bands 3 and 4. 
Information about their central frequencies, spectral resolution, and sensitivity, are given in Table~\ref{tab:setup}. 
For all the spectral windows, the sources used as flux and bandpass calibrators were J0423-0120 in band 3 and J0854+2006 in band 4, while J0547+2721 and J0550+2326 were used as gain (amplitude and phase) calibrators. 
The uncertainties in the flux calibration are $\sim 5\%$. 
The primary beam (i.e. the full width at half maximum of the main beam of a 12m antenna) ranges from $\sim 67$\asec\ at 84.5~GHz to $\sim 57$\asec\ at 99.3~GHz, and from $\sim 41$\asec\ at 140~GHz to $\sim 37$\asec\ at 153.5~GHz.

The calibrated data were produced by the calibration pipeline of 
the Common Astronomy Software Applications ({\sc CASA}, McMullin et al. 2007).
The pipeline version used is casapipe-6.4.1. 
Imaging and deconvolution were 
then performed on the calibrated uv tables with the {\sc gildas}\footnote{https://www.iram.fr/IRAMFR/GILDAS/}
software, after conversion of the calibrated uv tables from measurement sets in uvfits format and then in {\sc gildas} format. 
To be sensitive to the source extension, the clean maps were created using natural weighting. 
As significant emission is detected at the edge of the primary beam, we analyse the primary-beam-corrected images.
We attempted to perform self-calibration, but the self-calibrated images did not show any significant improvement due to the faintness of the continuum emission, and hence we analysed the non-self-calibrated images.
In the channel maps, the root mean square (rms) varies between 1.8 and 3.9 mJy beam$^{-1}$ in band 3, and between 8.8 and 18 mJy beam$^{-1}$ in band 4.
In the continuum images, the rms is $\sim 0.025$ mJy beam$^{-1}$ in band 3 and $\sim 0.1$ mJy beam$^{-1}$ in band 4. 
The angular resolution in both ALMA bands is $\sim 1.4-1.5$\asec, corresponding to
$\sim 0.1$~pc, or $\sim 20000$ au. 
In all the images, the maximum recoverable scale is $\sim 20$\asec.
The detected molecular lines and their spectroscopic parameters are listed in Table~\ref{tab:lines}.

\begin{table}
\begin{center}
\setlength{\tabcolsep}{1.8pt}
\caption{\label{tab:lines} Spectral parameters of detected lines.}
\begin{tabular}{l c c c}
\hline
\hline
 Quantum Numbers\tablefootmark{a}   &  Rest Freq.  & log[$A_{\rm ij}$]  &  $E_{\rm up}$  \\
               &   (GHz)            & (s$^{-1}$) &  (K)   \\
\hline
\multicolumn{4}{c}{\cyclic }   \\
$J(K_a,K_b)=2(1,2)-1(0,1)$ & 85.3389 & --4.6341 & 6.4 \\
\hline
\multicolumn{4}{c}{HCS$^+$}   \\
$J=2-1$ & 85.3479 & --4.9548 & 6.1 \\
\hline
\multicolumn{4}{c}{C$_4$H}  \\
        $N=9-8$,$J=19/2-17/2$ & & & \\
        $F=9-8$            & 85.6340 & --4.8189 & 20.5 \\
        $F=10-9$           & 85.6340 & --4.8163 & 20.5 \\
        $N=9-8$,$J=17/2-15/2$ & & & \\
        $F=8-7$            & 85.6726 & --4.8217 & 20.5 \\
        $F=9-8$            & 85.6726 & --4.8184 & 20.5 \\
\hline
\multicolumn{4}{c}{HCO}  \\
\hline
$N_{K_a,K_b}=1_{0,1}-0_{0,0}$ & & & \\
 $J=3/2-1/2$ & & & \\
 $F=2-1$                     & 86.6708 & --5.3289 & 4.2  \\
 $F=1-0$                     & 86.7084 & --5.3377 & 4.2  \\
 $F=1-1$                     & 86.7775 & --5.3366 & 4.2  \\
 $F=0-1$                     & 86.8058 & --5.3268 & 4.2  \\
 \hline
\multicolumn{4}{c}{HN$^{13}$C}  \\
$J=1-0$          & 87.0909 & --4.7288 & 4.2 \\
\hline
\multicolumn{4}{c}{\HCOpI }   \\  
$J=1-0$          & 86.7543 & --4.4142 & 4.2 \\
\hline
\multicolumn{4}{c}{\METH } \\
$J(K_a,K_b)=2(1,2)-1(1,1)$ E$_2$ & 96.7394 & --5.5923 & 12.5  \\
$J(K_a,K_b)=2(0,2)-1(0,1)$ A$^+$  & 96.7414  &  --5.4675 & 7.0  \\
$J(K_a,K_b)=2(0,2)-1(0,1)$ E$_1$  & 96.7445  &  --5.4676 & 20.1  \\
\hline
\multicolumn{4}{c}{$^{34}$SO}      \\
$J(K)=3(2)-2(1)$              & 97.7153 & --4.9695 & 9.1 \\
\hline
\multicolumn{4}{c}{CS}      \\
$J=2-1$              & 97.9810 & --4.7763 & 7.1 \\
\hline
\multicolumn{4}{c}{SO}      \\
$J(K)=3(2)-2(1)$     & 99.2999 & --4.9488 & 9.2 \\
\hline
\multicolumn{4}{c}{\FORM }  \\
$J(K_a,K_b)=2(1,2)-1(1,1)$ & 140.8395 & --4.2754 & 21.9\\
$J(K_a,K_b)=2(1,1)-1(1,0)$            & 150.4983 & --4.1890 & 22.6 \\
\hline
\end{tabular}
\end{center}
\tablefoot{
\tablefoottext{a}{All spectral parameters are taken from the Cologne Database for Molecular Spectroscopy (CDMS\footnote{https://cdms.astro.uni-koeln.de/cdms/portal/}; \citealt{endres16}), except those of HCO, which are taken from the Jet Propulsion Laboratory \citep[JPL,][]{pickett98}) catalogue;}
}
\end{table}

\section{Analysis of the maps}
\label{res}

\FloatBarrier
\begin{table*}[h!]
\begin{center}
\setlength{\tabcolsep}{6pt}
\caption{\label{tab:cont} Parameters of the millimeter continuum sources}
\begin{tabular}{l c c c c c c c c c c}
\hline
\hline
 Core &  R.A.   & Dec.    &  $F_{\nu,\rm 3mm}$\tablefootmark{a} &  $F_{\nu,\rm 2mm}$\tablefootmark{a} & $\theta_{\rm s}$\tablefootmark{b} & $D$\tablefootmark{c} & $N({\rm H_2})$\tablefootmark{d} & $M_{\rm d}$\tablefootmark{d} & $n({\rm H_2})$\tablefootmark{d} & $\beta$\tablefootmark{e} \\
        &  $h:m:s$ & $\circ:\prime:\prime\prime$ & Jy                   &   Jy                 & \asec\         & pc & $\times 10^{21}$\cmq\ &  $M_{\odot}$ & $\times 10^4$ \cmc\ &  \\
\hline
  N      & 05:37:41.55 & 36:07:31.9  &  $3.0\times 10^{-4}$  & $8.3 \times 10^{-4}$ & 2.4  & 0.18 & 5.7(1.2) & 2.9(1.5) & 1.1 & 0.34 \\
  S      & 05:37:42.31 & 36:07:15.2  &  $1.4\times 10^{-4}$  & $3.9 \times 10^{-4}$ & 2.0  & 0.15 &  2.1(0.5) & 0.7(0.4) & 0.5 & 0.21 \\
\hline
\end{tabular}
\end{center}
\tablefoot{
\tablefoottext{a}{Flux density integrated in the $3 \sigma$~rms level;}
\tablefoottext{b}{Deconvolved angular diameter;}
\tablefoottext{c}{Deconvolved linear diameter;}
\tablefoottext{d}{H$_2$ mass computed from the dust thermal continuum emission as described in Sect~\ref{fit-cores}, assuming a dust temperature equal to the excitation temperature of \METH\ (Table~\ref{tab:lines-cont}), and a dust mass opacity coefficient $\beta=1.7$;}
\tablefoottext{e}{Dust opacity index computed comparing the integrated flux densities at the two frequencies observed as described in Sect~\ref{fit-cores}.}
}
\end{table*}

\subsection{Millimeter continuum emission}
\label{maps-cont}

\begin{figure}
    \centering
    \includegraphics[width=1\hsize]{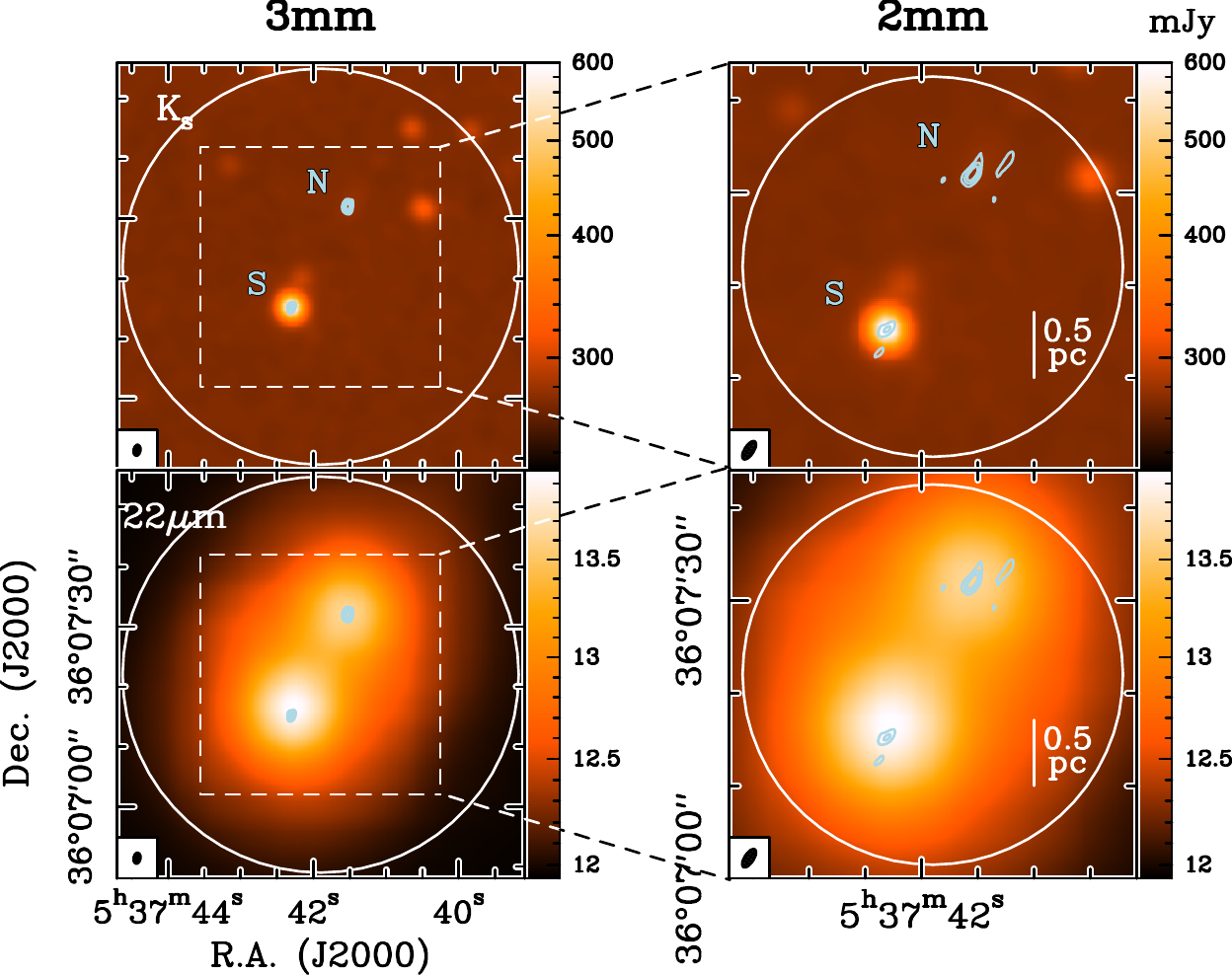}
    \caption{Continuum emission detected towards WB89-670 with ALMA. 
    The light blue contours in the left and right panels correspond to the 3~mm and 2~mm continuum emission, respectively. 
    Contours start from the 3$\sigma$ rms level ($\sim 8\times 10^{-5}$~Jy at 3~mm and $\sim 3 \times 10^{-4}$~Jy at 2~mm), and are in steps of $\sim 5 \times 10^{-5}$~Jy and $\sim 2 \times 10^{-4}$~Jy, respectively. 
    The synthesised beam is depicted in the bottom left corner, and the white circle indicates the ALMA primary beam at the two wavelengths. 
    The region shown in the 2~mm images corresponds to the dashed square illustrated in the top-left panel.
    The heat colour image in background is the $K_{\rm s}$ band of 2MASS (Fig.~\ref{fig:2mass}) in the top panels, and the WISE 22$\mu$m band in the bottom panels.}
    \label{fig:cont}
\end{figure}

The millimeter continuum emission detected towards WB670 in both ALMA bands is shown in Fig.~\ref{fig:cont}.
Two compact sources are detected. 
The strongest one, that we call N, is detected in the 3~mm image with signal-to-noise of 14 (peak flux density $\sim 3.5\times 10^{-4}$~Jy).
Core N is undetected or barely detected in the near-infrared, and it is clearly detected in the mid-infrared.
The other one, called S, coincides with the main infrared source detected in all 2MASS and WISE images, and is detected in the 3~mm image with signal-to-noise of 7 (peak flux density $\sim 1.8\times 10^{-4}$~Jy).
The flux densities, $F_{\nu}$, integrated inside the 3$\sigma$~rms contour levels are given in Table~\ref{tab:cont} for both sources, together with other observational properties.

\subsection{Spectra extracted from the millimeter continuum sources}
\label{spectra-cont}

\begin{figure*}
    \centering
    \includegraphics[width=0.9\hsize]{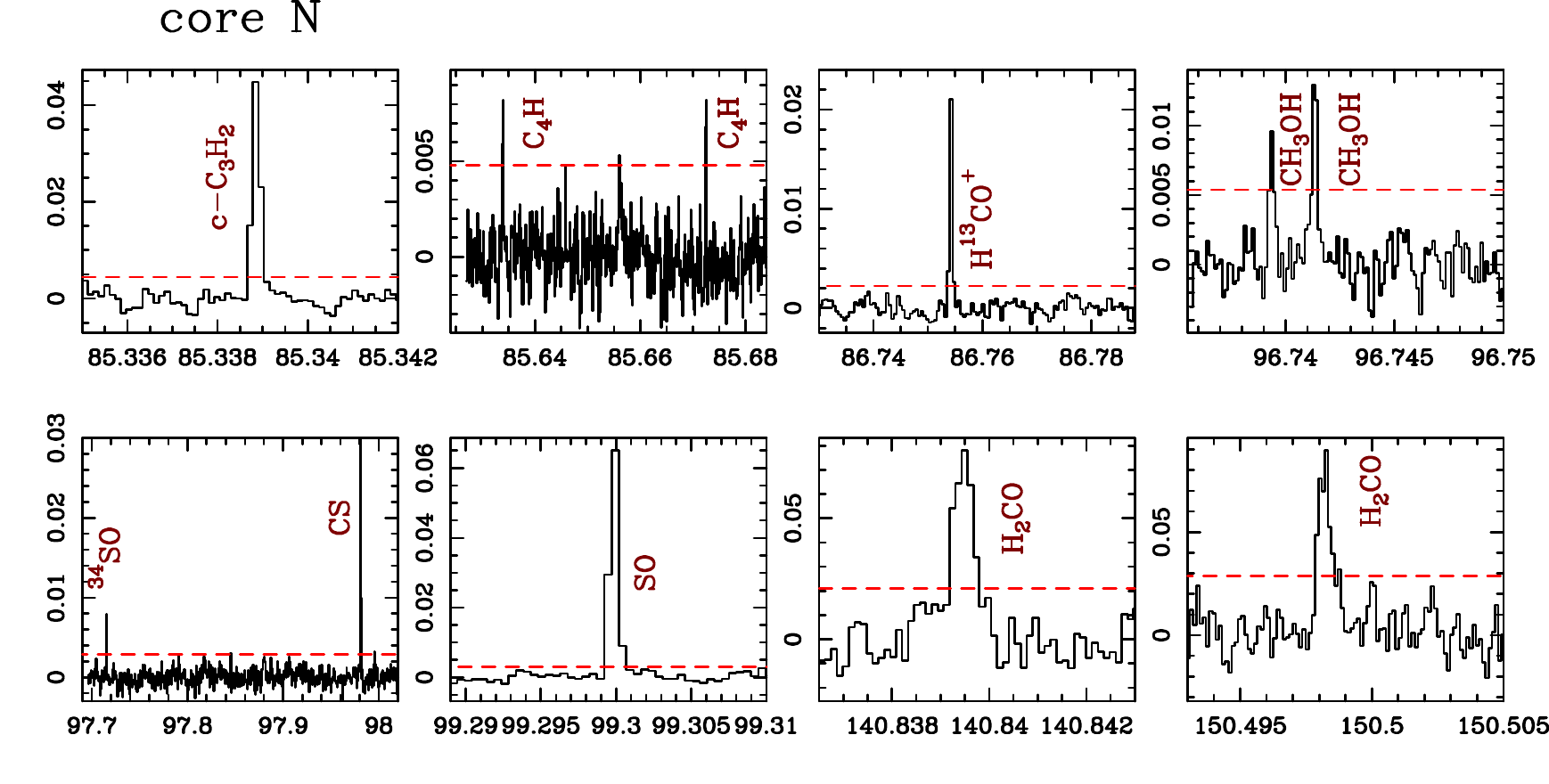}
    \includegraphics[width=0.9\hsize]{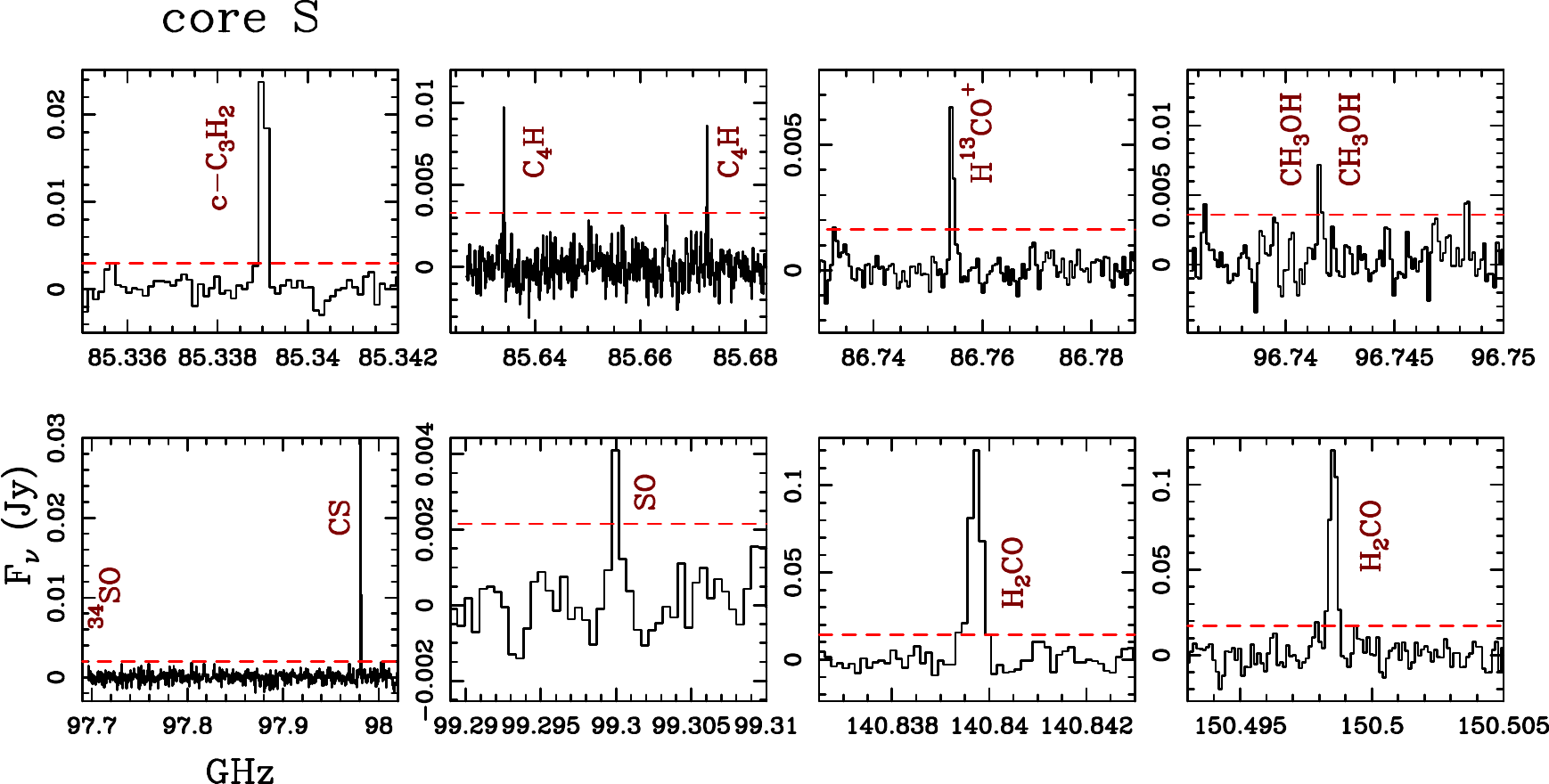}
    \caption{Spectra extracted from the 3$\sigma$~rms level contour of the 3~mm continuum for cores N (top) and S (bottom) (Fig.~\ref{fig:cont}).
    In the eight spectral windows, the clearly detected transitions are labelled, and the 3$\sigma$~rms level is represented by the red dashed line.
    The undetected species, namely HCS$^+$, HCO and HN$^{13}$C, are not shown.}
    \label{fig:spec-cont}
\end{figure*}

The integrated spectra extracted from the contour level at $3\sigma$~rms of the 3~mm continuum emission towards cores N and S (Fig.~\ref{fig:cont}) are shown in Fig.~\ref{fig:spec-cont}. 
The figure shows the detected transitions and the 3$\sigma$~rms level in the spectra.
The chemical richness towards N and S is similar, but the line intensities are overall higher towards N. 
We will discuss in more detail the chemical differences between the two cores in Sect.~\ref{analysis}.

\subsection{Emission morphology of molecular lines}
\label{maps-mol}

\begin{figure*}
    \centering
    \includegraphics[width=0.9\hsize]{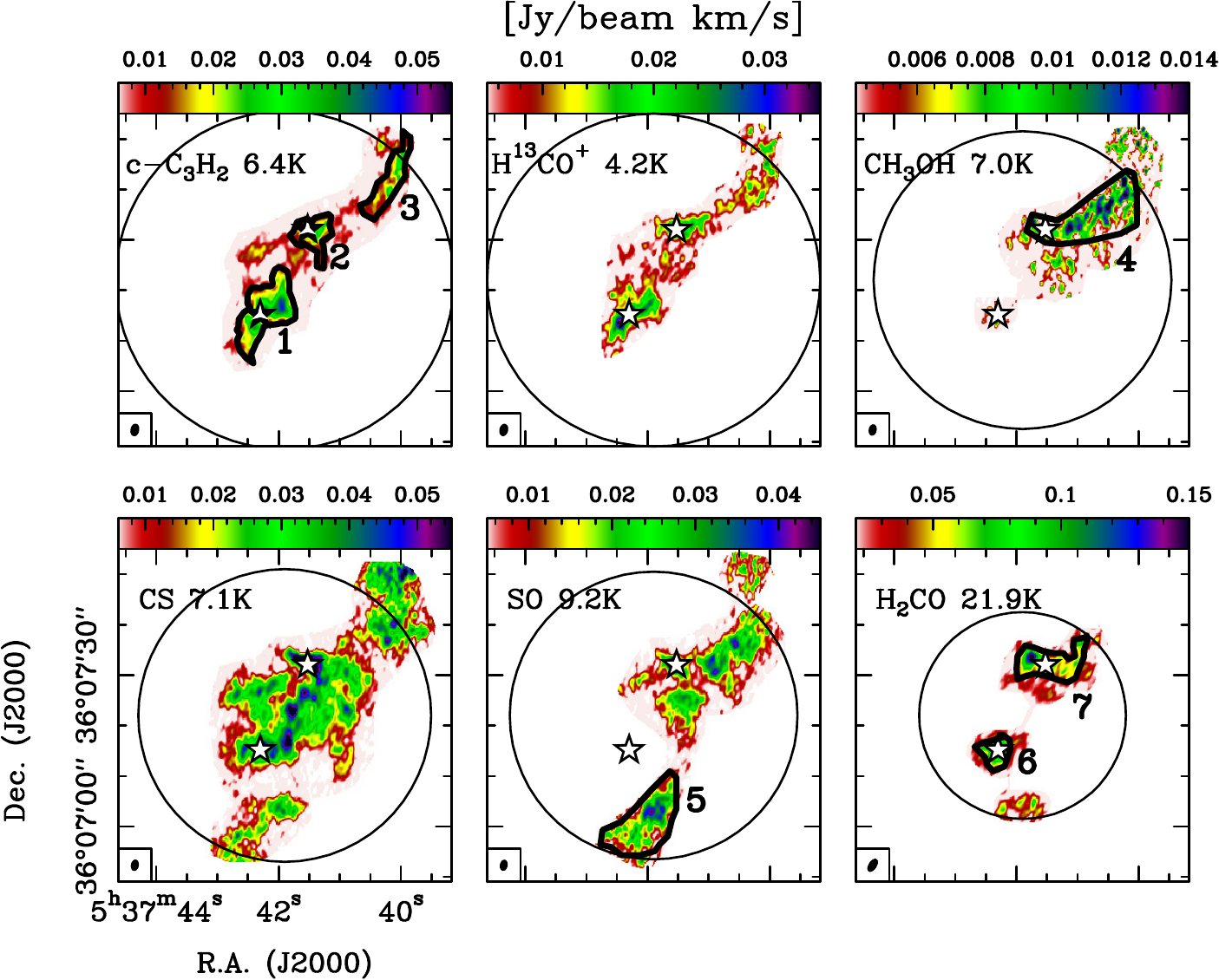}
    \caption{\label{fig:morphology} Velocity-integrated molecular emission maps.
    We show the six molecular species for which the signal-to-noise ratio is good.
    For C$_4$H, HCS$^+$, HCO, HN$^{13}$C, and $^{34}$SO, the maps are too noisy and we do not show them.
    The species, and the $E_{\rm up}$ of the line used in temperature units, are indicated in the top-left corner of each frame.
    The two stars indicate the peak position of the millimeter continuum cores N and S.
    The black circles highlight the primary beam at the rest frequency of each line (Table~\ref{tab:lines}).
    The integrated emission was computed in the channels with intensity higher than the 3$\sigma$~rms level.
    The velocity intervals used are: [$-18.4;-16.5$]~\kms\ for \cyclic; [$-19;-16$]~\kms\ for \HCOpI; [$-18.75;-16;25$]~\kms\ for \METH\ (from the line centred at $\sim 96.7414$~GHz) at [$-20;-14$]~\kms\ for CS; [$-19.5;-15.5$]~\kms\ for SO; [$-18.8;-16.2$]~\kms\ for \FORM\ (from the line centred at $\sim 140.8395$~GHz).
    The seven regions highlighted in black are those analysed in detail in Sect.~\ref{spectra-mol}, and correspond grossly to the 5$\sigma$~rms emission contour of the corresponding integrated intensity map (within the ALMA primary beam).}
\end{figure*}

The molecular transitions listed in Table~\ref{tab:lines} were detected in some regions of the final images with a signal-to-noise ratio $\geq 5$ (the 1$\sigma$ rms level is in Table~\ref{tab:setup}).
For each species, the velocity-integrated emission of the most intense transitions, namely \cyclic\ $J(K_a,K_b)=2(1,2)-1(0,1)$, \HCOpI\ $J=1-0$, \METH\ $J(K_a,K_b)=2(0,2)-1(0,1){\rm A}^+$, CS $J=2-1$, SO $J(K)=3(2)-2(1)$, and \FORM\ $J(K_a,K_b)=2(1,2)-1(1,1)$, is shown in Fig.~\ref{fig:morphology}. 
The integration interval in velocity is defined by the channels with signal-to-noise ratio $\geq 3$ (see caption of Fig.~\ref{fig:morphology}).
The other species detected, namely HCS$^+$, C$_4$H, HCO, HN$^{13}$C, and $^{34}$SO, are too faint across the whole mapped region to derive a good integrated map showing their morphology.

Overall, the emission peak of \cyclic, \HCOpI, and \FORM\ coincides with that of the dust continuum millimeter cores N and S, while that of SO and \METH\ is located towards a north-western elongated feature that extends for $\sim 20$\asec\ from core N to the north-western border of the primary beam.
The emission is extended in all tracers, and shows a filamentary structure oriented SE-NW, whose extension and width depends on the tracer. 
The filament is narrow in \cyclic, \HCOpI, and \METH, and goes roughly from the millimeter core S to the north-western edge of the primary beam. 
The emission of CS is much broader and arises also from a clump located to the south-eastern edge of the primary beam. 
This southern clump, extended about $\sim 15$\asec\ in NW-SE direction, is detected also in SO and \FORM, while is not detected in \cyclic, \HCOpI, and \METH.
The emission of SO resembles that of \METH\ in the northern part of the source and around core N, while towards the southern part of the source and around core S is different.
The emission of \FORM\ is compact and arises mostly from the two millimeter cores, maybe because the $E_{\rm u}$ of their transitions are higher ($\sim 22-23$~K) than those of the other lines ($\sim 4-12$~K).
The presence of \FORM\ in the northern filament cannot be determined because of the limited primary beam, but the integrated emission seems to follow the filament at least inside the primary beam.

To allow for a better inspection of the emission structure of each tracer, the six maps in Fig.~\ref{fig:morphology} are also shown, enlarged, in Figs.~\ref{fig:mean-1}--\ref{fig:mean-6} of Appendix~\ref{app:maps}.

\begin{table}[]
    \centering
        \caption{Angular and linear size of the seven molecular emitting regions shown in Fig.~\ref{fig:morphology}.}
    \begin{tabular}{ccc}
\hline 
\hline
Region   & $\theta_{\rm s}$  &  $D_{\rm s}$ \\
         & (\asec )          &   (pc)           \\
\hline
 \multicolumn{3}{c}{\cyclic}  \\
\hline   
   1      & 11.6 & 0.85 \\
   2      & 7.5 & 0.55 \\
   3      & 8.3 & 0.61 \\
\hline
 \multicolumn{3}{c}{\METH}  \\
\hline   
   4      & 14.5 & 1.06 \\
   \hline
 \multicolumn{3}{c}{SO}  \\
\hline   
   5      & 12.9 & 0.94\\
\hline
 \multicolumn{3}{c}{\FORM}  \\
\hline   
   6      & 6.1 & 0.45 \\
   7      & 9.5 & 0.69 \\
\hline
    \end{tabular}
    \label{tab:size}
\end{table}

\subsection{Spectra extracted from molecular emission regions}
\label{spectra-mol}

The complexity of the source structure described in Sect.~\ref{maps-mol} and the fact that different species emit preferentially in different volumes indicates a chemical differentiation at small spatial scales.
Due to this differentiation, we analyse the molecular emission dividing the source in seven regions, based on the emission maps of the species with highest differentiation.
The regions are shown in Fig.~\ref{fig:morphology}, and are defined roughly on the integrated emission maps of \cyclic, \METH, SO, and \FORM: three are based on the dominant clumps detected in \cyclic\ (labelled as 1, 2, and 3 in Fig.~\ref{fig:morphology}), one is based on the \METH\ emission in the northwestern filament (4), and one on the southern clump seen in SO (5).
To complement the analysis, we also extract spectra from the two dominant clumps detected in \FORM\ (6 and 7), which overlap partially with regions 1 and 2 though.

The spectra extracted from these seven regions are shown in Figs.~\ref{fig:spec-C3H2} and \ref{fig:spec-H2CO}.
In Table~\ref{tab:size}, we give the equivalent angular and linear size of the seven regions, derived as the diameter of a circle with the same area.
The regions have sizes in between 0.45 and 1.06~pc.

\subsection{Comparison between ALMA and IRAM spectra}
\label{resolved}

The \cyclic, \HCOpI, and C$_4$H lines in Table~\ref{tab:lines} were observed also with the IRAM-30m telescope by \citet{fontani22a}.
To evaluate if extended emission was resolved out by the interferometer, we extracted the integrated ALMA spectra in flux density units of the mentioned lines from a circular region with size equal to that of the IRAM 30m telescope main beam ($\sim 27$\asec).
Then, we converted the IRAM 30m spectra from main beam temperature units to flux density units through the equation \citep[see e.g.][]{fontani21}:
\begin{equation}
  F_{\nu}{\rm [mJy]} = \frac{T_{\rm MB}{\rm [K]}}{1222}{\nu{\rm [GHz]}^2 \Theta[\prime \prime]^2}  \;,
  \label{eq:conversion}
\end{equation}
where $F_{\nu}$ is the flux density at the observed rest frequency $\nu$, $T_{\rm MB}$ is the main beam temperature, and $\Theta$ the main beam angular size of the IRAM 30m telescope at frequency $\nu$.
\begin{figure}
    \centering
    \includegraphics[width=0.95\hsize]{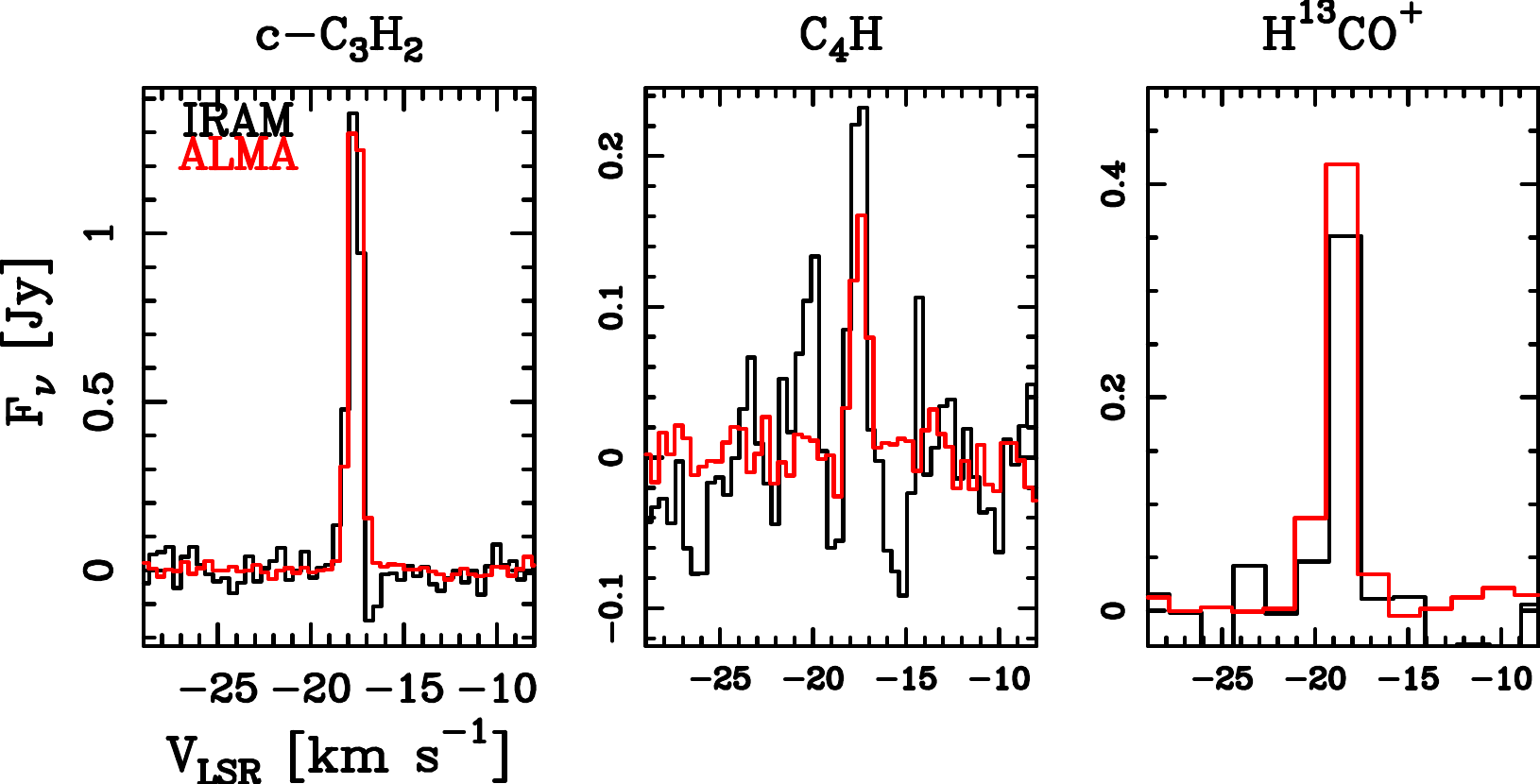}
    \caption{Flux density comparison between ALMA and IRAM 30m spectra.
    The IRAM 30m and ALMA spectra are shown as black and red histograms, respectively.
    The ALMA spectra were extracted from an angular region equivalent to the IRAM 30m beam at the frequency of the lines ($\sim 27$\asec).}
    \label{fig:flux-comparison}
\end{figure}

The comparison is shown in Fig.~\ref{fig:flux-comparison}.
The flux densities measured with ALMA are consistent with those measured with the IRAM-30m telescope in \cyclic\ and \HCOpI\ within a factor $\leq 10\%$, fully consistent with the calibration errors.
For C$_4$H, we show in Fig.~\ref{fig:flux-comparison} the $J=19/2-17/2$ hyperfine component, for which the difference between the IRAM and ALMA spectra is the largest (a factor $\sim 1.3-1.4$).
Considering the calibration errors and the rms noise in the spectra (in particular in the C$_4$H IRAM 30m spectrum), we conclude that the amount of extended flux resolved out by ALMA is negligible and does not affect significantly our analysis.

\section{Spectral analysis and derivation of column densities}
\label{coldens}

\subsection{Spectral analysis}
\label{analysis}

Flux densities were converted to brightness temperatures (\Tb) according to Eq.~\ref{eq:conversion}.
This relation needs the observed angular equivalent diameter (i.e. the diameter of the equivalent circle) of each core. 
Therefore, $T_{\rm B}$ is an average brightness temperature over the solid angle of each core.
We have analysed the spectra in $T_{\rm B}$ units with the MAdrid Data CUBe Analysis ({\sc madcuba}\footnote{{\sc madcuba} is a software developed in the Madrid Center of Astrobiology (INTA-CSIC) which enables to visualise and analyse single
spectra and data cubes. https://cab.inta-csic.es/madcuba/},
Mart\'in et al.~\citeyear{martin19}) software.
The transitions reported in Table~\ref{tab:lines} were  modelled via the Spectral Line
Identification and LTE Modelling (SLIM) tool of {\sc madcuba}. The lines
were fitted with the AUTOFIT function of SLIM. 
This function produces the synthetic spectrum that best matches the data
assuming a constant excitation temperature (\Tex) for all transitions of the same species. 
The other input parameters are: total molecular column
density (\Ntot), radial systemic velocity of the source ($V$), line
full-width at half-maximum (FWHM), and angular size of the
emission ($\theta_{\rm s}$). 
AUTOFIT assumes that $V$, FWHM, and \thetaS\ are
the same for all transitions. 
For \thetaS, we assumed that the emission is more extended than the extraction region in all tracers, and hence no filling factor is appliyed in the derivation of the best fit parameters.
Assuming as \thetaS\ the equivalent radii given in Table~\ref{tab:size} would change the resulting column densities by at most a factor 1.2.
We left all other parameters free except \Tex, that we had to fix for all species except \METH, which is the only molecule for which two transitions with significantly different $E_{\rm u}$ were detected (Table~\ref{tab:lines}).
We decided to adopt for all species the \Tex\ measured from \METH.

The analysis described above with {\sc madcuba} was adopted for the lines of these species: \cyclic, \METH, \FORM, C$_4$H, and HCS$^+$.
For the remaining ones, namely HCO, \HCOpI, HN$^{13}$C, CS, SO, and $^{34}$SO, {\sc madcuba} cannot fit the lines in all regions because the observed FWHMs are comparable to the spectral resolution of their spectra (spw 33, 39, and 41 in Band 3).
Towards region 5, \METH\ could also not be fit for the same reason.
Therefore, for these species, we obtained the integrated intensities of the lines with the {\sc class}\footnote{https://www.iram.fr/IRAMFR/GILDAS/doc/pdf/class.pdf} package of the {\sc gildas} software.
\Ntot\ were then calculated with the same equations used by {\sc madcuba}, namely assuming LTE at \Tex\ equal to that derived from \METH\ also in this case.
We assumed also optically thin conditions.
The assumption is justified both by the Gaussian line profiles, and by the fact that the optical depths provided by AUTOFIT for the lines analysed with {\sc madcuba} are consistent with optically thin conditions.

The results obtained are listed in Table~\ref{tab:lines-cont}.
One remarkable immediate result is the small values of the FWHMs of all lines, which are narrower than $\sim 2$~\kms\ both in N and S.
Even though the cores harbour infrared-bright sources, the FWHMs are narrow and comparable to those measured in starless cores in infrared dark clouds \citep[e.g.][]{kong17,barnes23} rather than to those measured in infrared-bright protostellar envelopes \citep[e.g.][]{fontani02,fontani23}.
The excitation temperatures derived from \METH\ are $\sim 9$~K and $\sim 15$~K in core N and S, respectively.
Such low temperatures could indicate sub-thermal excitation conditions, at least for \METH. 
Assuming the same \Tex\ for the other molecules, as explained above, 
we derived \Ntot\ of the order of $\sim 10^{12}$~\cmq\ for \HCOpI, and of the order of $\sim 10^{13}$~\cmq\ for \METH, C$_4$H, \FORM, CS, and SO.

As stated above, to derive \Ntot\ we fixed \Tex\ to that of \METH\ for all molecules.
This assumption is justified by the fact that all transitions analysed have energies of the upper level in between $\sim 4$ and $\sim 20$~K, similar to the energy range of the methanol lines (7--20~K).
However, the computed low \Tex\ could indicate sub-thermal conditions for \METH.
We estimate the uncertainty introduced by this simplified approach varying \Tex\ in between the \METH\ values and a representative kinetic temperature of the gaseous envelope of high-mass protostellar objects $\sim 30$~K \citep[e.g.][]{sanchez13,fontani15}, in case \METH\ emission is sub-thermally excited.
The \Ntot\ variation is dependent of the species and it is on average larger for core N, for which the difference between \Tex\ from \METH\ (9~K) and the representative temperature of 30~K is the highest.
For example: in core N, \Ntot\ of CS increases from 1.7$\times 10^{13}$ to 2.6$\times 10^{13}$~\cmq;
\Ntot\ of \HCOpI\ increases from 1.5$\times 10^{12}$ to 2.8$\times 10^{12}$~\cmq;
\cyclic\ cannot be reasonably fit with a \Tex\ above 15~K, if one considers both the detected transition and the upper limit on the transition $J(K_a,K,b)=4(3,2)-4(2,3)$, at $\sim 85.656$~GHz but undetected;
\Ntot\ of C$_4$H decreases from 9.1$\times 10^{12}$ to 5.5$\times 10^{12}$~\cmq;
\Ntot\ of \FORM\ remains the same within the errors.

\subsection{Fractional abundances in the continuum cores}
\label{fit-cores}

To estimate the molecular fractional abundances with respect to H$_2$, we compute the column density of H$_2$, $N({\rm H_2})$, from the integrated continuum flux density through the equation \citep[e.g.][]{battersby14}:
\begin{equation}
 N({\rm H_2}) = \frac{\gamma F_{\nu}}{\Omega \kappa_{\nu} B_{\nu}(T_{\rm d})\mu ({\rm H_2}) m_{\rm H}}\;, 
\end{equation}
 where $\gamma$ is the gas-to-dust mass ratio, $F_{\nu}$ is the flux density at frequency $\nu$, $\Omega$ is the source solid angle at the observed frequency, $\kappa_{\nu}$ is the dust opacity usually parametrised as $\kappa_{\nu}=\kappa_{\nu_0}(\nu/nu_0)^{\beta}$ \citep{oeh94}, $B_{\nu}(T_{\rm d})$ is the Planck function at dust temperature $T_{\rm d}$, $\mu({\rm H_2})$ is the mean molecular weight for which we will adopt 2.8 \citep{kauffmann08}, and $m_{\rm H}$ is the mass of the hydrogen atom. 
Using the empirical relation between $\gamma$ and $R_{\rm GC}$ derived from \citet{giannetti17}, we find $\gamma \sim 3000$ at $R_{\rm GC}=23.4$~kpc.
Considering the uncertainties in the \citet{giannetti17} empirical relation, $\gamma$ is $3000^{+700}_{-2200}$, hence more than 8 times higher than the conventional value of 100.
We adopt for $T_{\rm d}$ the excitation temperature derived from \METH\ (Table~\ref{tab:lines-cont}) for each core, assuming coupling between gas and dust.
We also assume dust opacity index $\beta =1.7$, a typical value used for ice-coated dust grains in dense cores, and $\kappa_{\nu_0}=0.899$ at $\nu_0=230$~GHz \citep{oeh94}.
The resulting $N({\rm H_2})$ are $\sim 5.7 \times 10^{21}$ and $\sim 2.1 \times 10^{21}$~\cmq, respectively.
Assuming spherical sources, we also compute the core H$_2$ gas masses, M$_{\rm d}$, and volume densities, $n({\rm H_2})$. 
M$_{\rm d}$ and $n({\rm H_2})$ are of the order of $\sim 1$~\solm\ and of $\sim 10^{4}$~\cmc, respectively, for both cores, with core N being more dense and massive.
All parameters are listed in Table~\ref{tab:cont}.
We show the molecular fractional abundances derived from the \Ntot-to-$N({\rm H_2})$ ratio in Table~\ref{tab:lines-cont}.

These estimates are affected by several uncertainties. First, as discussed in Sect.~\ref{analysis}, \Ntot\ can vary up to a factor 2 depending on the species.
Second,
computing $\beta$ from the integrated continuum flux densities in the 3~mm and 2~mm bands (Tab.~\ref{tab:cont}) through Eq.(2) in \citet{chacon19b}, we find $\beta \sim 0.34$ and $\beta \sim 0.21$ for core N and S, respectively.
With these $\beta$, we find $N{\rm (H_2)}\sim 1.6\times 10^{21}$~\cmq\ for N and $N{\rm (H_2)}\sim 1.6\times 10^{21}$~\cmq\ for S, respectively, i.e. a factor $\sim 3$ smaller than those given in Table~\ref{tab:cont}.
The molecular abundances in Table~\ref{tab:lines-cont}, hence, would increase systematically by the same factor.
Another big uncertainty is introduced by the value of $\gamma=3000^{+700}_{-2200}$ calculated from \citet{giannetti17}, which makes the abundances to vary up to an additional factor $\sim 4$.
There are also the uncertainties on $\beta$, which depend on the measured continuum flux densities and dust temperatures.
While the relative error on the continuum flux densities is $\sim 10\%$, that on the dust temperature is difficult to quantify.
However, we stress that all these uncertainties affect the individual abundances but not their ratios.

In the diffuse ISM, $\beta$ is $\sim 2$ \citep[e.g.][]{del84}, while
lower values are measured in dense cores \citep[e.g.][]{forbrich15,galametz19} and circumstellar discs \citep[e.g.][]{testi14, friesen18}. 
$\beta$ can depend a lot on grain size, composition, and porosity, but values lower than 0.5 are hard to explain without considering large grains of size $\sim 100$~$\mu$m -- 1~mm \citep[e.g.][]{testi14,ysard19}.
In protoplanetary discs, such large grains are expected as a result of coagulation and growth.
However, cores N and S have a linear diameter $\geq 30000$~au, hence the observed emission is unlikely dominated by a circumstellar disc.
Values of $\beta$ lower than 1 were measured also up to $\sim 2000$~au scales around young protostars \citep[e.g.][]{galametz19}, indicative of early grain growth, although theory is struggling in finding the possible reasons for such growth at the relatively low density of the $1000-10000$~au scale of protostellar envelopes.
It has been proposed that large grains are not formed on such extended envelope scales, but transported there from the site of growth via jets and winds \citep[e.g.][]{cacciapuoti24}.
This scenario could be possible for WB670, but the narrow FWHMs of the observed lines in N and S (Tab.~\ref{tab:lines-cont}) suggest a very quiescent environment, and hence should be disregarded.
\citet{silsbee22} proposed that the presence of very small grains could also cause a low $\beta$.
This scenario may be possible if the distribution of grain size at the Galactocentric distance of WB670 is different from that in the local medium.
Other options are also possible, like high dust opacities, and contamination from free-free emission at 3 mm. 
We did not find in the literature any free-free emission study towards WB670 (e.g. from radio-continuum emission), and hence this option cannot be checked.

\FloatBarrier
\begin{table}[h!]
\begin{center}
\setlength{\tabcolsep}{1.8pt}
\caption{\label{tab:lines-cont} Best-fit parameters of the millimeter continuum sources N and S spectra.}
\begin{tabular}{c c c c c c}
\hline
\hline
  molecule   & $V$   &  FWHM &  \Tex\ & \Ntot\ & [X]\tablefootmark{a} \\
     & \kms\ & \kms\   &   K    & $\times 10^{12}$\cmq\   & $\times 10^{-9}$ \\
\hline
\multicolumn{6}{c}{N} \\
 \cyclic\ & --17.40(0.01) & 0.74(0.02) &  & 7.9(0.2) & 1.4 \\
 HCS$^+$ & -- & -- &  & $\leq 0.4$ & $\leq 0.07$ \\
 C$_4$H & --17.32(0.05) & 0.9(0.1) &  & 19(2) & 3.3 \\
 HCO & -- & -- &  & $\leq 3.2$ & $\leq 0.6$ \\
 HN$^{13}$C & -- & -- &  & $\leq 0.3$ & $\leq 0.05$ \\
 H$^{13}$CO$^+$ & --17.37(0.07) & 2.01(0.08) &  & 1.5(0.1) & 0.3 \\
 CH$_3$OH & --17.4(0.1) & 0.92(0.08) & 9(2) & 2.2(0.3) & 0.4 \\
 CS & --17.46(0.03) & 2.00(0.04) &    & 17(2) & 3.0 \\
 SO & --17.44(0.02) & 2.01(0.05) &     & 43(1)  & 7.5 \\
 \FORM\ & --17.37(0.04) & 1.25(0.09) &  & 8.5(0.5) & 1.5 \\
  \hline
  \multicolumn{6}{c}{S} \\
  \hline
 \cyclic\ & --18.0(0.1) & 0.7(0.1) &  & 7.6(0.8) & 3.6 \\
 HCS$^+$ & -- & -- &  & $\leq 0.6$ & $\leq 0.3$ \\
 C$_4$H & --17.93(0.03) & 0.69(0.06) &  & 44(4) & 21 \\
 HCO   & -- & -- &  & $\leq 2.1$ & $\leq 1$ \\
 HN$^{13}$C & -- & -- &  & $\leq 0.4$ & $\leq 0.2$ \\
 H$^{13}$CO$^+$ & --18.0(0.6) & 1.6(1) &  & 1.0(0.3) & 0.5 \\
 CH$_3$OH & --18.0(0.7) & 1.1(0.5) & 15(3) & 23(3) & 11 \\
 CS & --17.92(0.02) & 1.87(0.03) &  & 11.7(1) & 5.6 \\
 SO & --17.8(0.2) & 2.1(0.4) &  & 3.4(0.8) & 1.6 \\
 \FORM\ & --17.96(0.01) & 0.63(0.03) & & 8.5(0.8) & 4.0 \\ 
\hline
\end{tabular}
\end{center}
\tablefoot{
\tablefoottext{a}{Fractional abundance with respect to H$_2$, computed as \Ntot/$N({\rm H_2})$. 
$N({\rm H_2})$ is calculated from the continuum flux density in Table~\ref{tab:cont} as explained in Sect.~\ref{fit-cores}.}
}
\end{table}

\subsection{Column densities in the molecular emitting regions}
\label{fit-molecules}

In Appendix~\ref{app:fits}, we list the parameters obtained fitting the lines with {\sc madcuba} and {\sc class} towards the lines detected in the seven regions indicated in Fig.~\ref{fig:morphology}.
As for the continuum cores, the FWHMs are always narrower than 1.5-2~\kms, indicating quiescent gas.
The \Tex\ derived from the two detected \METH\ lines are in between 6.4~K and 15~K, similar to that measured towards cores N and S, suggesting that the emission arise from cold gas also in these regions.
This is consistent again with the relatively low FWHMs of the lines, smaller than $\sim 1-2$~\kms, which indicates further that the emission arise from quiescent material.
Regarding \Ntot, for \METH\ we derive values of the order of $10^{12}-10^{13}$\cmq, with the maximum value ($\sim 1.6 \times 10^{13}$\cmq) in the north-western filament in between core N and the edge of the primary beam.
The hydrocarbons C$_4$H and \cyclic\ have similar column densities, both of the order of $\sim 10^{12}$\cmq.
For these regions we cannot derive abundances because we cannot estimate the H$_2$ column densities, hence we will discuss the comparison between the \Ntot\ of the various species in Sect.~\ref{comparison}.
    
\section{Discussion}
\label{discu}

\subsection{Column density comparisons}
\label{comparison}

\begin{figure*}
    \centering
    \includegraphics[width=0.9\hsize]{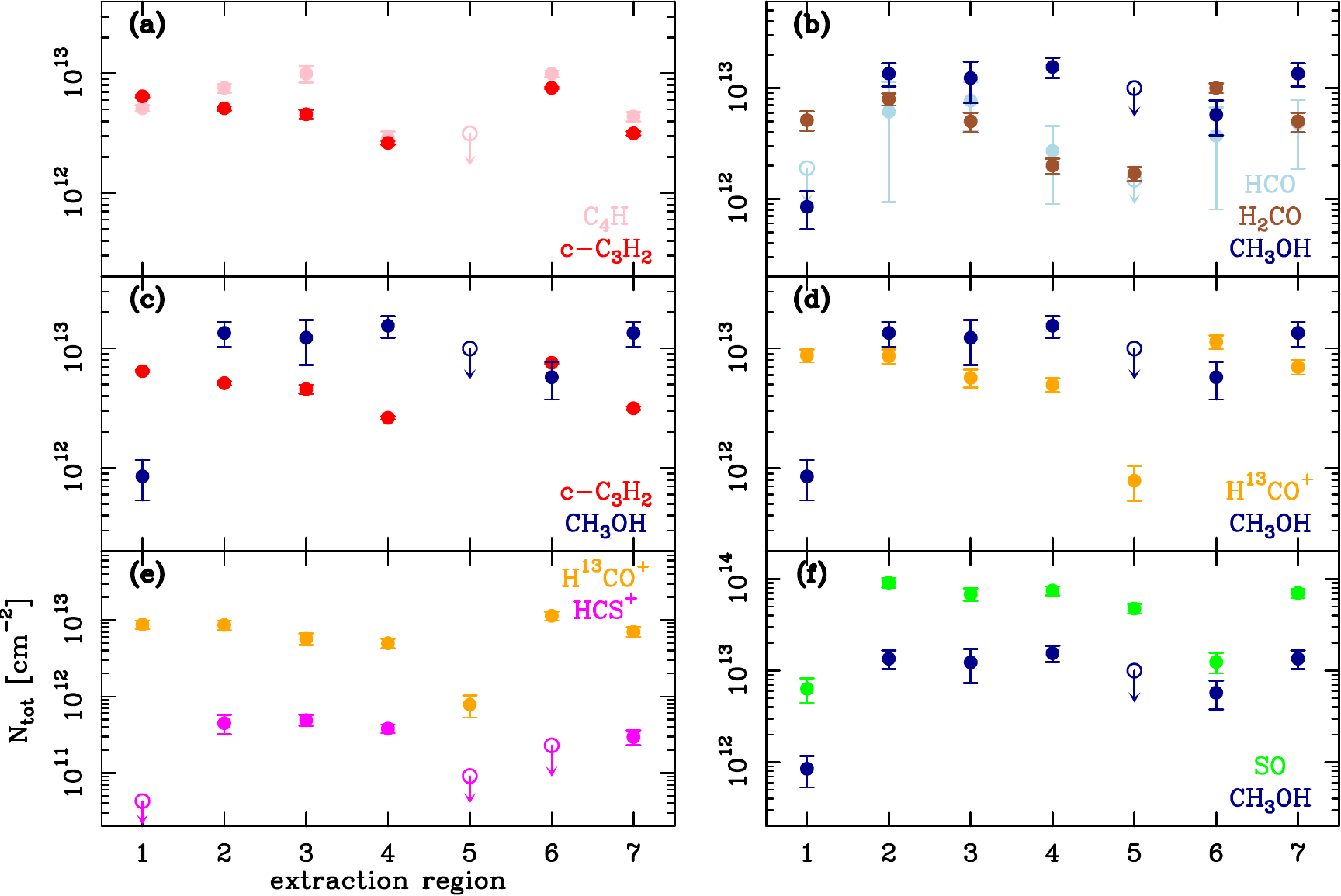}
    \caption{Column density comparison in the seven molecular extraction regions. 
    The colours in the different panels indicate the different molecular species according to the labels in the bottom-right corner (top-right corner in panel (e)).
    Empty symbols with an arrow pointing downwards are upper limits.}
    \label{fig:coldens-comparisons}
\end{figure*}

In Fig.~\ref{fig:coldens-comparisons} we compare the molecular column densities between different species in the seven regions of WB670.
The relative \Ntot\ of the hydrocarbons C$_4$H and \cyclic\ are similar and do not change significantly in the sub-regions of WB670, as shown in panel (a) of Fig.~\ref{fig:coldens-comparisons}.
Both species require atomic C not locked in CO to form, therefore their good agreement in the different regions is consistent with the expectations.
In panel (b) of Fig.~\ref{fig:coldens-comparisons} we compare \Ntot\ of \FORM, HCO, and \METH,
which are thought to be chemically related because all can be formed from hydrogenation of CO on dust grains. 
\Ntot\ of \METH\ is largely variable in the seven sub-regions, and the trend does not follow that of HCO and \FORM, since the higher \Ntot\ of \METH\ are found where those of HCO and \FORM\ are lower, and vice versa.

The same dichotomy is apparent also between \METH\ and \cyclic\ (panel (c)), especially evident in region 1, where \Ntot\ of \METH\ is the lowest and that of \cyclic\ is the highest, and in region 4, where the opposite happens.
A similar dichotomy is observed in the pre-stellar core L1544 \citep[][]{spezzano16,jensen23}, where the emission of the two molecules seems anti-correlated.
In particular, \METH\ emission arises from colder and more shielded regions of the L1544 core envelope, while \cyclic\ emission overlaps with the dust continuum emission.
Such observational difference is consistent with the different physical conditions needed to form the two molecules.
While it is well-known that \METH\ is formed from sequential hydrogenation of CO on grain surfaces \citep[e.g.][]{fuchs09},
the formation of \cyclic\ occurs in the gas-phase through an ion-molecule reaction followed by dissociative recombination \citep[e.g.][]{sipila16}.
These reactions do not need a low-temperature and high-density environment like that required for \METH\ formation.
Therefore, the dichotomy between these two species likely arises from such different formation routes.
Interestingly, the variation of \Ntot\ in the seven regions found for \FORM\ and HCO (panel (b) in Fig.~\ref{fig:coldens-comparisons}) resembles more that of \cyclic\ than that of \METH.
This points to formation routes of \FORM\ and HCO in gas-phase rather than from surface chemistry processes.
In particular, while \METH\ can be formed only on the surfaces of dust grains given the inefficiency of gas phase routes at
low temperatures \citep[e.g.][]{garrod06}, \FORM\ is also known to form in the gas phase from regions rich in hydrocarbons, where C is not yet completely locked in CO \citep[see e.g.][]{chacon19}.
HCO can also form via gas-phase reactions \citep{rivilla19}.
Our observations seem to indicate that the gas-phase formation of \FORM\ and HCO from atomic C is likely more efficient than its formation on dust grains, despite the lower abundance of C at such large Galactocentric distances.

The trends of the molecular ions \HCOpI\ and HCS$^+$ are also more in line with hydrocarbons and \FORM\ rather than with \METH\ (panels (d) and (e)).
Again, this can be attributed to a formation of these ions only in gas-phase.
The species for which the trend in \Ntot\ is the most similar to \METH\ is SO, as indicated in panel (f) of Fig.~\ref{fig:coldens-comparisons}. 
This similarity could be explained by their common origin from surface chemistry.
In fact, SO is thought to be formed in gas phase from atomic S \citep[e.g.][]{vidal17}, more abundant upon grain sputtering than in the diffuse gas.

\subsection{Comparison with local and inner Galaxy star-forming regions}
\label{local}

Extrapolating the elemental abundance trends measured by \citet{mendez22} at the Galactocentric distance of WB670, the oxygen, carbon, and nitrogen fractional abundances should be: [O/H]$\sim 6.7\pm 2.3\times 10^{-5}$, [C/H]$\sim 1.8\pm 0.5\times 10^{-5}$, and [N/H]$\sim 5.3 \pm 1.3\times 10^{-6}$, respectively.
The reference values at the Solar circle are: [O/H]$\sim 3.1\pm 1.0\times 10^{-4}$, [C/H]$\sim 2.6\pm 0.8\times 10^{-4}$, and [N/H]$\sim 4.7 \pm1.4 \times 10^{-5}$, respectively
\citep{mendez22}.
Therefore, the relative elemental ratios [O/C] and [O/N] should increase from [O/C]$\sim 1.2\pm 0.7$ and [O/N]$\sim 6.6 \pm 4.0$ to [O/C]$\sim 3.7\pm 2.0$ and [O/N]$\sim 12.6 \pm 7.6$.
A molecular ratio that, in principle, can be particularly sensitive to changes in the [O/C] ratio is \METH/\cyclic\ as found in the pre-stellar core L1544 \citep[e.g.][]{spezzano16}, since to form \METH\ one needs carbon locked in CO and to form \cyclic\ one needs free atomic carbon.
\begin{figure}
    \centering
    \includegraphics[width=0.95\hsize]{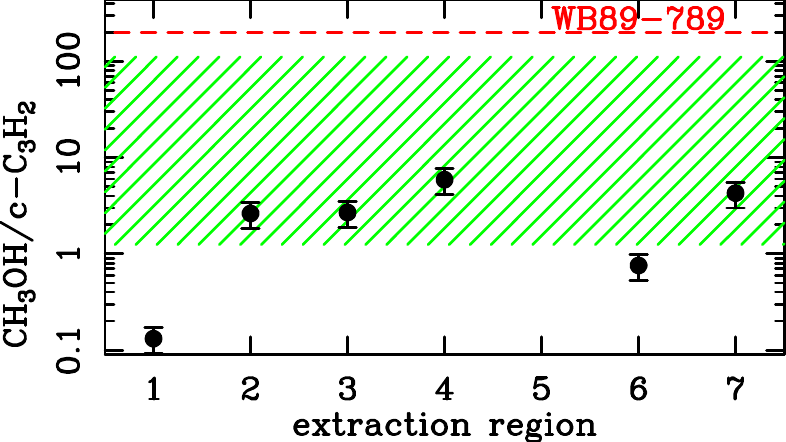}
    \caption{Column density ratio \METH/\cyclic\ in WB670 and other star-forming regions.
    The black points indicate the ratios measured in the seven regions of WB670 (with the exception of region 5, for which both column density estimates are upper limits).
    The green area correspond to the range of values measured in local star-forming regions \citep{higuchi18}, and the red dashed line is the ratio measured in the outer Galaxy hot-core WB789 \citep{shimonishi21}.}
    \label{fig:c3h2-ch3oh}
\end{figure}
Figure~\ref{fig:c3h2-ch3oh} shows the column density ratios \METH/\cyclic\ in the seven molecular regions of WB670, and compare them to the ratios measured in local (low- and high-mass) star-forming regions \citep{higuchi18} and in the outer Galaxy hot-core WB89-789 \citep[hereafter WB789,][]{shimonishi21}.
As representatives of local low-mass star-forming regions, we took the cores in the Perseus molecular clouds observed by \citet{higuchi18}.
As representatives of high-mass star-forming regions, we took IRAS 20126 and AFGL 2591 \citep{freeman23}.
The linear scale resolved is $\sim 5000-10000$~au in \citet{higuchi18}, while it is $\sim 200000$~au in \citep{freeman23}.
The \METH/\cyclic\ ratios derived in WB670 agree on average with the lower values measured in local star-forming regions,
while the ratio found in WB789 is twice the highest local values (Fig.~\ref{fig:c3h2-ch3oh}).
Because WB789 is located in between the local Galaxy and WB670, there does not seem to be a monothonic trend of the \METH/\cyclic\ ratio with the Galactocentric distance.
We will better discuss the comparison between WB670 and WB789 in Sect.~\ref{WB789}.

\begin{figure*}
    \centering
    \includegraphics[width=0.9\hsize]{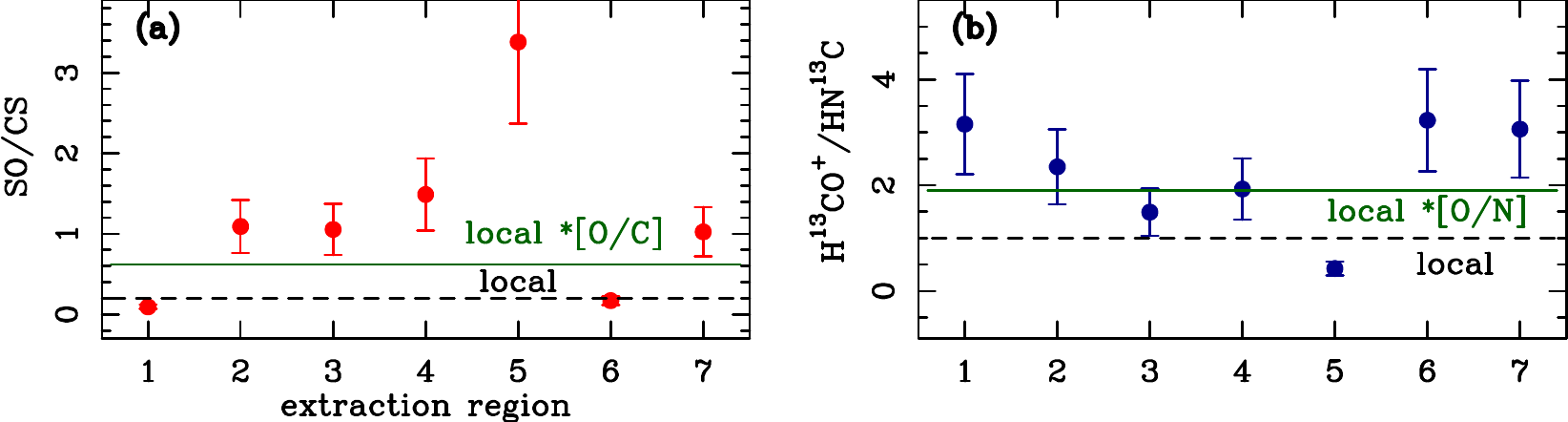}
    \caption{Column density ratios in the seven molecular extraction regions of WB670.
    Panel (a) shows the measured SO/CS ratios (points), and compare them to the average SO/CS ratio measured in a sample of inner and local Galaxy high-mass star-forming regions \citep[dashed line]{fontani23}, and to the
    expected SO/CS ratio obtained multiplying the local measured value (0.2, dashed line) by the increase in the elemental ratio [O/C]$\sim 3.1$ (horizontal green line) from the Solar circle to $R_{\rm GC}=23.4$~kpc \citep{mendez22}.
    Panel (b) shows the same comparison between the measured \HCOpI/HN$^{13}$C ratio and the expected ratio obtained multiplying the local measured value for the increase in the [O/N] elemental ratio.}
    \label{fig:elemental}
\end{figure*}

We also checked if the column density ratios we derive in WB670 between species that differ from just one element such as HN$^{13}$C and \HCOpI, and CS and SO, can eventually be attributed to the change in elemental ratios with $R_{\rm GC}$.
Figure~\ref{fig:elemental} shows the column density ratios SO/CS (panel (a)).
Despite a dispersion of an order of magnitude, the average value is SO/CS$\sim 1.2$.
\citet{fontani23} found average SO/$^{13}$CS ratios $\sim 11-12$ in a sample of high-mass star-forming cores in different evolutionary stages, and in the local and inner Galaxy.
This ratio translates into $\sim 0.2$ when considering a conversion factor $^{12}$C/$^{13}$C$\sim 68$ for the local ISM \citep{milam05} (dashed line in panel (a) of Fig.~\ref{fig:elemental}).
Scaling this value for the increase in the [O/C] elemental ratio ($\sim 3.1$) from the Solar circle to $R_{\rm GC}\sim 23.4$~kpc \citep{mendez22}, the expected [SO/CS] column density ratio in WB670 should be $\sim 0.62$, that is a factor 2 smaller than the average measured one.
But the disperion is such that, overall, the measured ratios are consistent with those measured in the Solar neighbourhoods scaled for metallicity.
Panel (b) of Fig.~\ref{fig:elemental} shows the \HCOpI/HN$^{13}$C column density ratios measured in WB670.
The average value is $\sim 2.6$.
In this case the dispersion is only a factor $\sim 2$.
In local and inner Galaxy star-forming regions this molecular ratio is very variable.
\citet{vasyunina11} measured an average \HCOp/HNC ratio of $\sim 10$ in infrared-dark clouds. 
\citet{zinchenko09} found \HCOpI/HN$^{13}$C$\sim 1$ in a sample of high-mass star-forming regions, indicating that this ratio can be very sensitive to changes in the physical conditions.
Because WB670 contains infrared-bright objects, its physical conditions are likely similar to those of the active star-forming regions observed by \citet{zinchenko09}. 
This means that the observed \HCOpI/HN$^{13}$C column density ratio increases on average by a factor 2.6 from local star-forming regions to $R_{\rm GC}\sim 23.4$~kpc, while the expected increase of the [O/N] elemental ratio should be $\sim 1.8$.
Therefore, again the increased elemental abundance ratio cannot fully explain the observed molecular ratios.
However, considering the dispersion of the values in the literature, and the error introduced by the assumed \Tex\ in our analysis, overall the ratios are consistent with a metallicity-scaled trend.
On the other hand, the elemental ratios beyond $R_{\rm GC}\sim 15$~kpc are poorly constrained by observations \citep[e.g.][]{romano20} and suffer from uncertainties up to $\sim 30\%$ on the individual elemental abundances \citep[][]{mendez22}, and hence their extrapolation to such large $R_{\rm GC}$ could be inaccurate.

\subsection{Comparison with sub-Solar metallicity hot-cores}
\label{WB789}

\begin{figure}
    \centering
    \includegraphics[width=0.9\hsize]{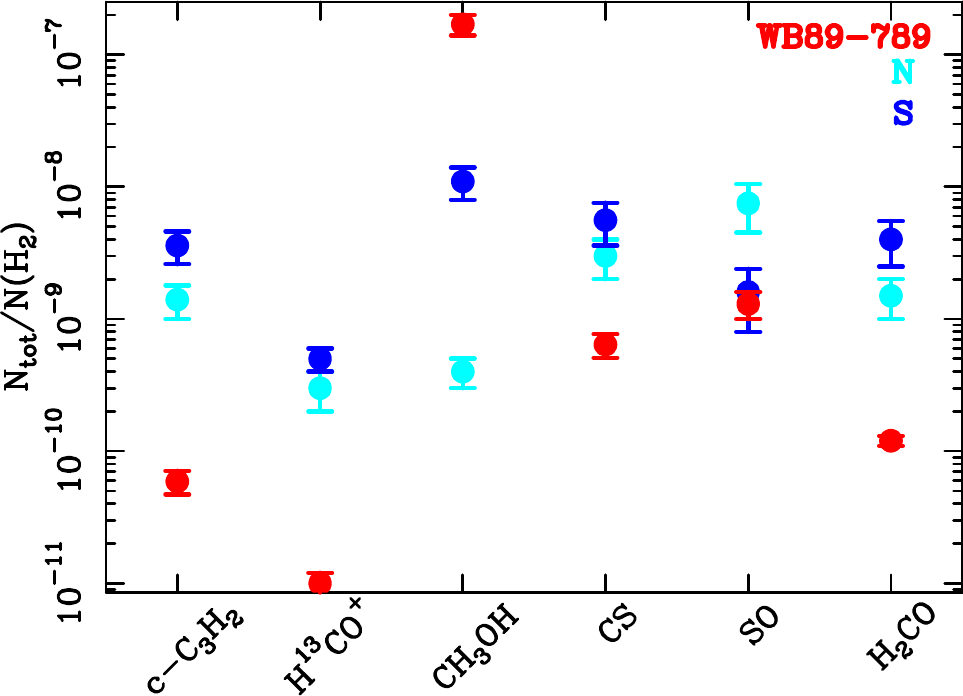}
    \caption{\label{fig:shimonishi} Fractional abundances with respect to H$_2$ derived for N, S, and WB789.
    The WB789 abundances were derived on a hot-core linear size of 0.1~pc \citep{shimonishi21}.}
\end{figure}

Some species detected in WB670 were also detected in the hot core WB789 \citep{shimonishi21}, located at $R_{\rm GC}\sim 19$~kpc.
The species in common are: \cyclic, \METH, \FORM, \HCOpI, SO, and CS.
\citet{shimonishi21} derive two abundance values, computed on linear diameters of 0.026 and 0.1~pc, respectively.
For the comparison with N and S, we use the 0.1~pc values because this size is more similar to that of N and S (Table~\ref{tab:cont}).
Figure~\ref{fig:shimonishi} shows the comparison: except for SO, all other species show abundances significantly different from each other, indicating a clear different chemical composition.
In particular, the \METH\ abundance in WB789 is two orders of magnitude higher than in N and S, while the abundances of all other carbon-bearing species are lower.
In fact, the \METH\ abundance in WB789 is $2 \times 10^{-7}$, while in N and S is $0.4\times 10^{-9}$ and $11\times 10^{-9}$, respectively.
Even considering the abundances obtained assuming $\beta$ in Table~\ref{tab:cont}, that are a factor $\sim 3$ higher than those listed in Table~\ref{tab:lines-cont}, the \METH\ abundances in N, S, and WB789 are not consistent.
\citet{shimonishi21} proposed a chemical stratification of the environment of the WB789 hot-core, with \METH\ and all COMs arising from the inner ($\leq 0.015$~pc) warm ($\geq 100$~K) region, and SO, CS, \FORM, \cyclic, \HCOpI, and HN$^{13}$C all associated with an external ($\geq 0.05$~pc) cold ($\leq 40$~K) envelope.
Our findings also indicate that the species we detect are associated with relatively cold and quiescent envelope material, including \METH.
However, the line widths in WB789 are always larger than $\sim 2$~\kms\ even in the envelope tracers, indicating that WB670 and WB789 are different kind of star-forming regions.
Simulations show that the abundance of \METH\ in dense star-forming core is extremely sensitive to dust temperature variations \citep{aeh15,peg18}.
In particular, the \METH\ production on grain surfaces depends on the duration of the early cold phase in which CO is hydrogenated on grain mantles, owing to the high volatility of atomic hydrogen \citep[see also][]{shimonishi20}.

The \METH\ abundances in N and S are comparable to the so-called organic-poor hot-cores detected in the Large Magellanic Cloud \citep[LMC,][]{shimonishi20,hamedani24}, characterised by values that cannot simply be explained scaling the abundances measured in the local Galaxy for the decreased metallicity \citep[about a factor 2-3, e.g.][]{andrievsky01,rolleston02} of the LMC.
\citet{shimonishi20} proposed that these organic-poor hot-cores in the LMC could be due to dust temperatures higher than those in organic-rich hot-cores.
Therefore, the lower \METH\ abundance in WB670 could be due to an inefficient (or insufficient) hydrogenation of CO in the stage of ice formation, due to either a (too) warm dust temperature, or to a (too) short cold ice formation stage, or to a (too) low gas density.
All scenarios would also predict a lower production efficiency of \FORM\ on ice mantles, and this would be in agreement with our finding that \FORM\ should be formed in gas phase in WB670.
Alternatively, WB670 can be in a late(r) evolutionary stage with respect to WB789, when the molecules in the inner protostellar cocoon(s), including those evaporated from dust grain mantles, have been mostly dissociated by the strong protostellar radiation field.
\citet{peg18} proposed that \METH\ abundance could be even enhanced in low-metallicity environments, owing to the lower C/O ratio which would imply that most of C is locked in CO, needed to form \METH.
However, this scenario would predict that the \METH\ abundance in WB670 should be higher than in WB789, because the C/O ratio in WB670 is lower than in WB789 (according to the elemental trends with $R_{\rm GC}$, see Sect.~\ref{intro}).
This is clearly at odds with our observational results.
We will better discuss the influence of metallicity on the observed molecular abundances in Sect.~\ref{modelling}.

\subsection{Chemical modelling}
\label{modelling}

To better investigate  if the measured column density ratios can constrain the initial elemental abundances, as well as other physical parameters that cannot be derived from the data, we compare our observational results to the prediction of chemical models. 
We investigate these clouds with the open-source gas-grain chemistry code UCLCHEM \citep{holdship17}. 
This time-dependent astrochemical code allows us to model the evolution of each molecule. 
We considered a "static model", namely an isothermal
cloud at constant density with a radius of $R=0.5$~pc (the maximum radius of the modelled regions), and we explored a range of physical and chemical conditions. 
The parameters include H$_2$ number density ($\mathrm{cm}^{-3}$), gas temperature, cosmic-ray ionisation rate $\zeta$, radiation field $F_{\rm UV}$, and both the initial oxygen and carbon abundances as listed in Table~\ref{tab:sobol_grid}.
Each of the models is ran with an edge visual extinction of $A_v=1.0$~mag, and molecular ratios are computed at $10^6$ years.
The grid is discussed in more detail in Vermari\"{e}n et al.~(in prep.). 
We converted between $^{12}$C$/^{13}$C using a conversion
factor of $10^{1.8}$ ($\sim 68$) since the chemical network did not include isotopologues.
For the modelling, we used a conventional gas-to-dust mass ratio of 100. 
We discuss the details of using the observed value obtained from \citet{giannetti17} at the end of this section.

\begin{figure*}
    \centering
    \includegraphics[width=\linewidth]{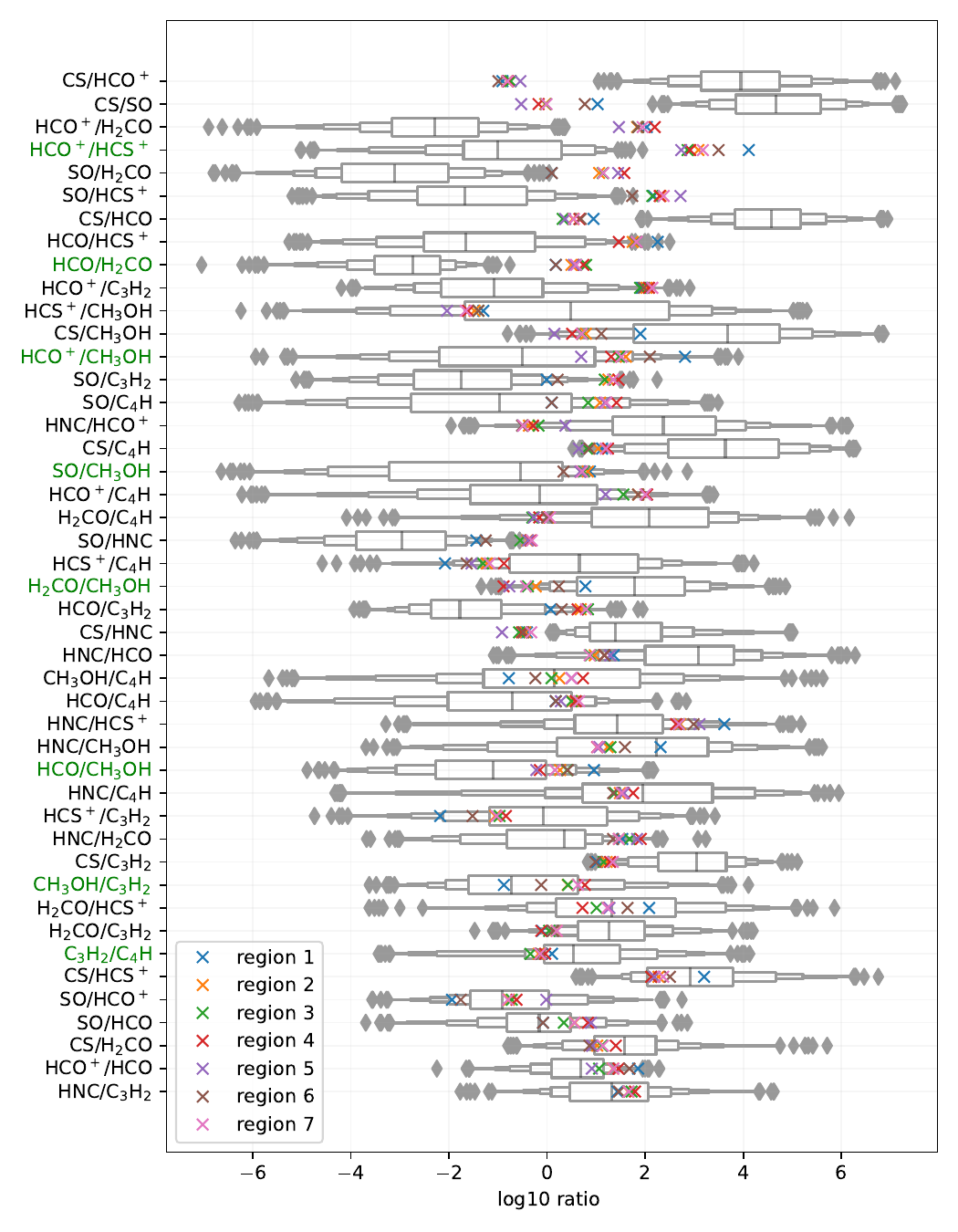}
    \caption{A comparison of the distribution of log-ratios of the observed species, with the ratios shown in Fig.~ \ref{fig:coldens-comparisons} highlighted in green.   The seven observed region's ratios (coloured markers) are plotted against the distribution of the modelled ratios (grey boxen plots). 
    The molecules are sorted from top to bottom in order of decreasing error.}
    \label{fig:all_ratios}
\end{figure*}

We excluded models for which the predicted fractional abundances of our molecules is below $10^{-13}$. 
This results in a total of 710 models out of the original 65536. 
These models can then be compared to the observations, and the distributions of the fit of the theoretical ratios with the observed ones are shown
in Fig.~\ref{fig:all_ratios}, sorted in descending averaged mean squared error between the models and 
observations. 
Some ratios, namely CS/HCO$^+$, CS/SO,  HCO$^+$/H$_2$CO,  HCO$^+$/HCS$^+$, SO/H$_2$CO, SO/HCS$^+$,
CS/HCO, HCO/H$_2$CO and CS/HNC, have no overlap between the models and the observations. 
Some observed ratios also lie in the tails of the model distribution.
Overall, it is very unlikely to find a model that fits all ratios well.

\begin{table*}[h]
    \centering
    \begin{tabular}{|c|c|c|c|}
    \hline
    Parameter & Min & Max & Sample space \\
    \hline
    Density $n_\mathrm{H}\;(\mathrm{cm}^{-3})$ & $1 \times 10^3$ & $1 \times 10^7$ & log \\
    Temperature $T\;(\mathrm{K})$& 10 & 100 & linear \\
    Cosmic-ray ionisation rate $\zeta\;(s^{-1})$& $1 \times 10^{-17}$ & $1 \times 10^{-14}$ & log \\
    Radiation field $F_{\mathrm{UV}}\;(\mathrm{Habing})$ & 0.1 & 100 & log \\
    Initial elemental abundance of oxygen $f_\mathrm{O}/f_\mathrm{O,\odot}$  & $0.05 \times 3.34 \times 10^{-4}$ & $1.0 \times 3.34 \times 10^{-4}$ & linear \\
    Initial elemental abundance of carbon $f_\mathrm{C}/f_\mathrm{C,\odot}$  & $0.05 \times 1.77 \times 10^{-4}$ & $1.0 \times 1.77 \times 10^{-4}$ & linear \\
    \hline
    \end{tabular}
    \caption{\label{tab:sobol_grid} Ranges of physical and chemical parameters of the grid of models.}
\end{table*}

\begin{figure*}
    \centering
    \includegraphics[width=\linewidth]{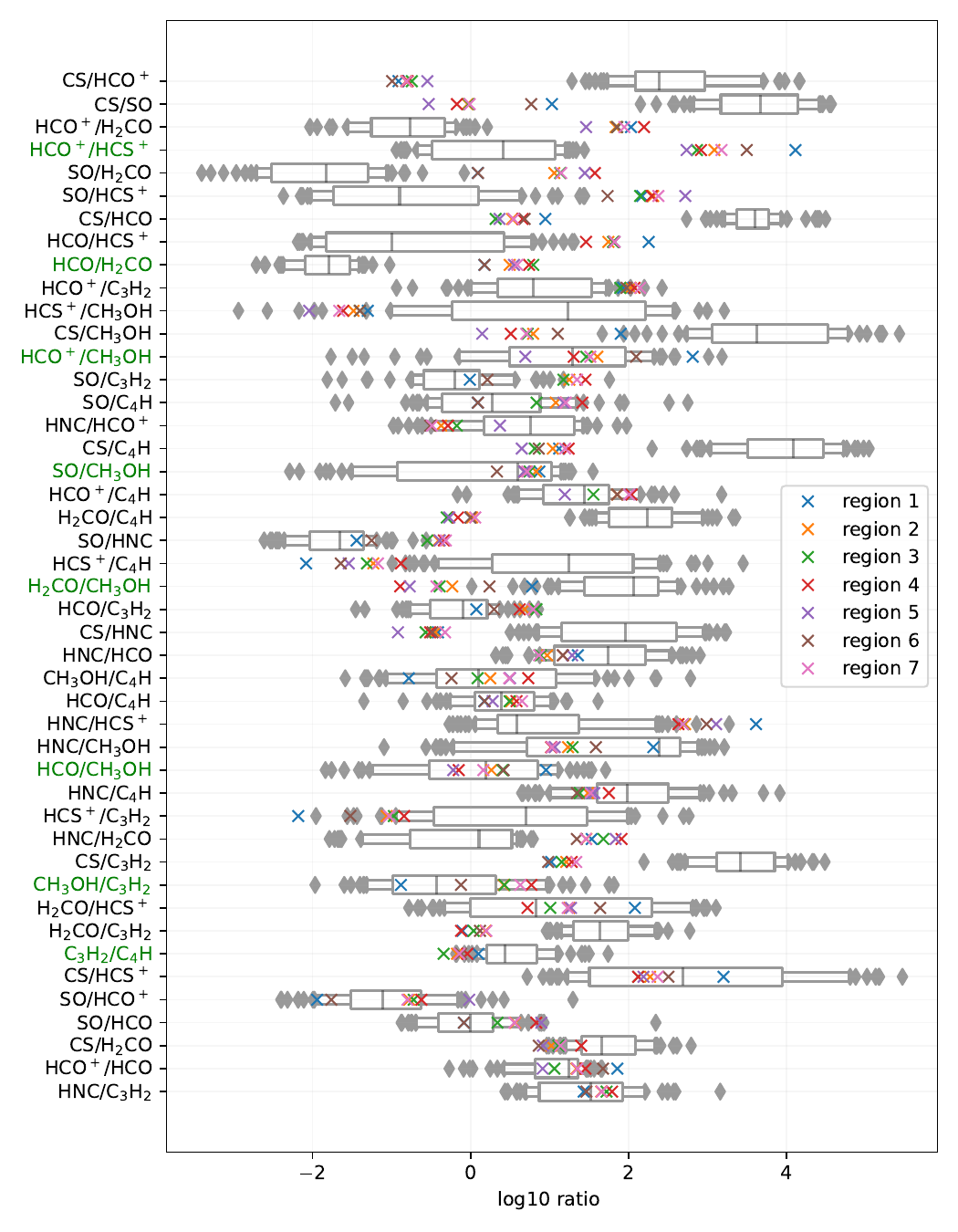}
    \caption{The best 50 best models (grey boxen plots) compared to the 7 observed regions (coloured markers). }
    \label{fig:filtered_ratios}
\end{figure*}

In order to further constrain the physical parameter space, the 50 models with the lowest Mean Squared Error (MSE) are chosen.
Figure~\ref{fig:filtered_ratios} shows again the comparison between the models and the observations.
The distribution
of the models is now closer to the observations, so the
horizontal axis is more compact than in Fig.~\ref{fig:all_ratios}.
Many of the ratios distance between models and observations decrease considerably by taking only the best models, but
the CS, HCS$^+$ and H$_2$CO based ratios do not improve.
 
\begin{figure}
    \centering
    \includegraphics[width=\linewidth]{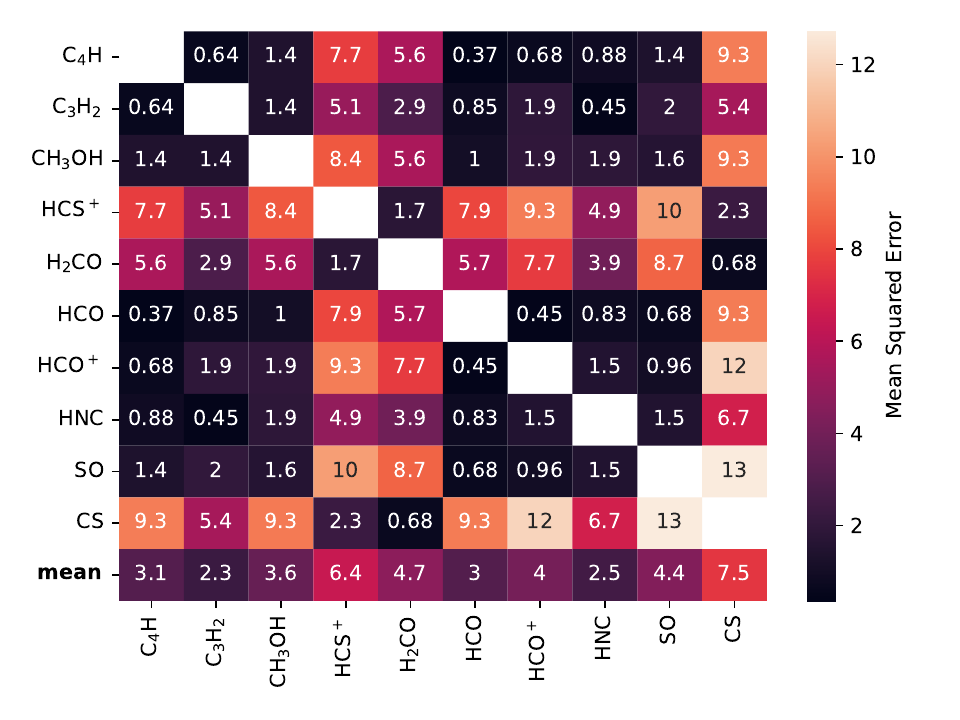}
    \caption{The mean squared error (MSE) computed between all of the regions and the best 50 models. 
    The averaged MSE for each molecule is included on the bottom row.}
    \label{fig:filtered_distances}
\end{figure}

\begin{figure}
    \centering
    \includegraphics[width=\linewidth]{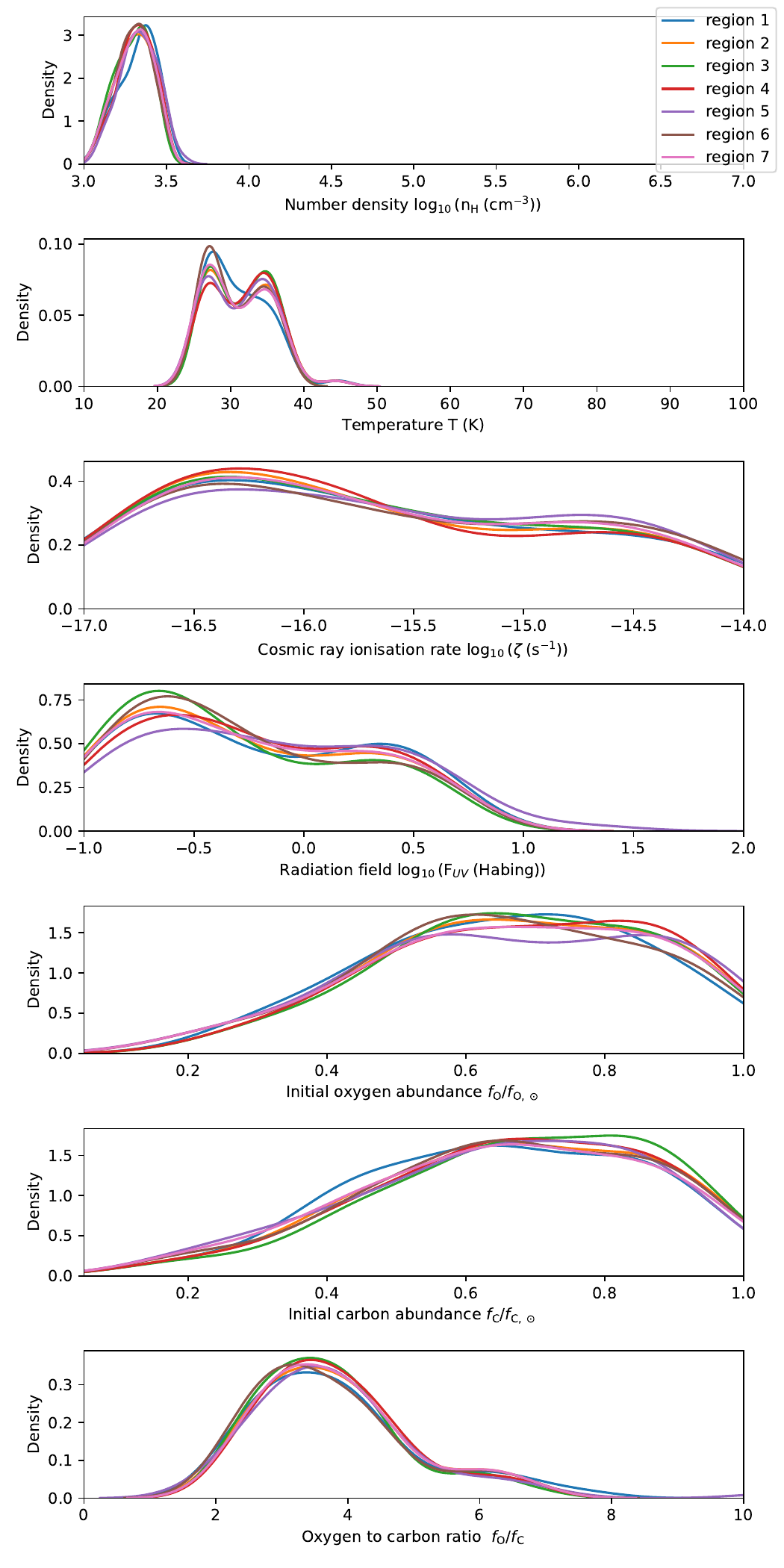}
    \caption{A kernel density probability estimate of the physical parameters for the best 50 models, fit separately for each of the regions, indicated by the coloured curves as labelled in the top-right.}
    \label{fig:filtered_parameters}
\end{figure}

In addition to the ratios that could not be explained by just the ``observable'' ratios, some ratios also have no matching distribution with the subset of best models. 
It does show however that we can fit
many of the ratios well with the subset of best models.
By plotting the error for each of the ratios in a pairwise grid, as can be seen in Fig.~\ref{fig:filtered_distances},
we can better understand which molecules
have the lowest error. 
This shows the molecules were best fit in the following order:
\cyclic, HNC, HCO, C$_4$H, \METH, HCO$^+$, SO, H$_2$CO, HCS$^+$, and finally CS. 
We then plot the distribution of the physical and chemical parameters investigated in the grid (Fig.~\ref{fig:filtered_parameters}). 
The models indicate densities
ranging from $10^3$ to $10^{3.6}$~\cmc, temperatures from 20 to 45~K, radiation fields lower than 5 Habing, and cosmic-ray ionisation rate $\zeta$ in a wide range from the interstellar value of $\sim 10^{-17}$~s$^{-1}$ up to $\sim 10^{-14}$~s$^{-1}$. 
Oxygen and carbon initial abundances both larger than 1/5th of the Solar value are favoured by the models. 
Additionally, the [O/C] ratio favours values in between $\sim 2$ and $5$,
with a small distribution of even higher oxygen enhancements.
The densities measured in N and S are higher ($\sim 10^4$~\cmc) than the values predicted by the models, which, however, refer to the molecular regions, all more extended than N and S.
Moreover, assuming $\beta$ as in Table~\ref{tab:cont} to compute $N{\rm (H_2)}$, we derive densities of the order of $\sim 10^3$~\cmc, consistent with model predictions.

\begin{figure*}
    \centering
    \includegraphics[width=\linewidth]{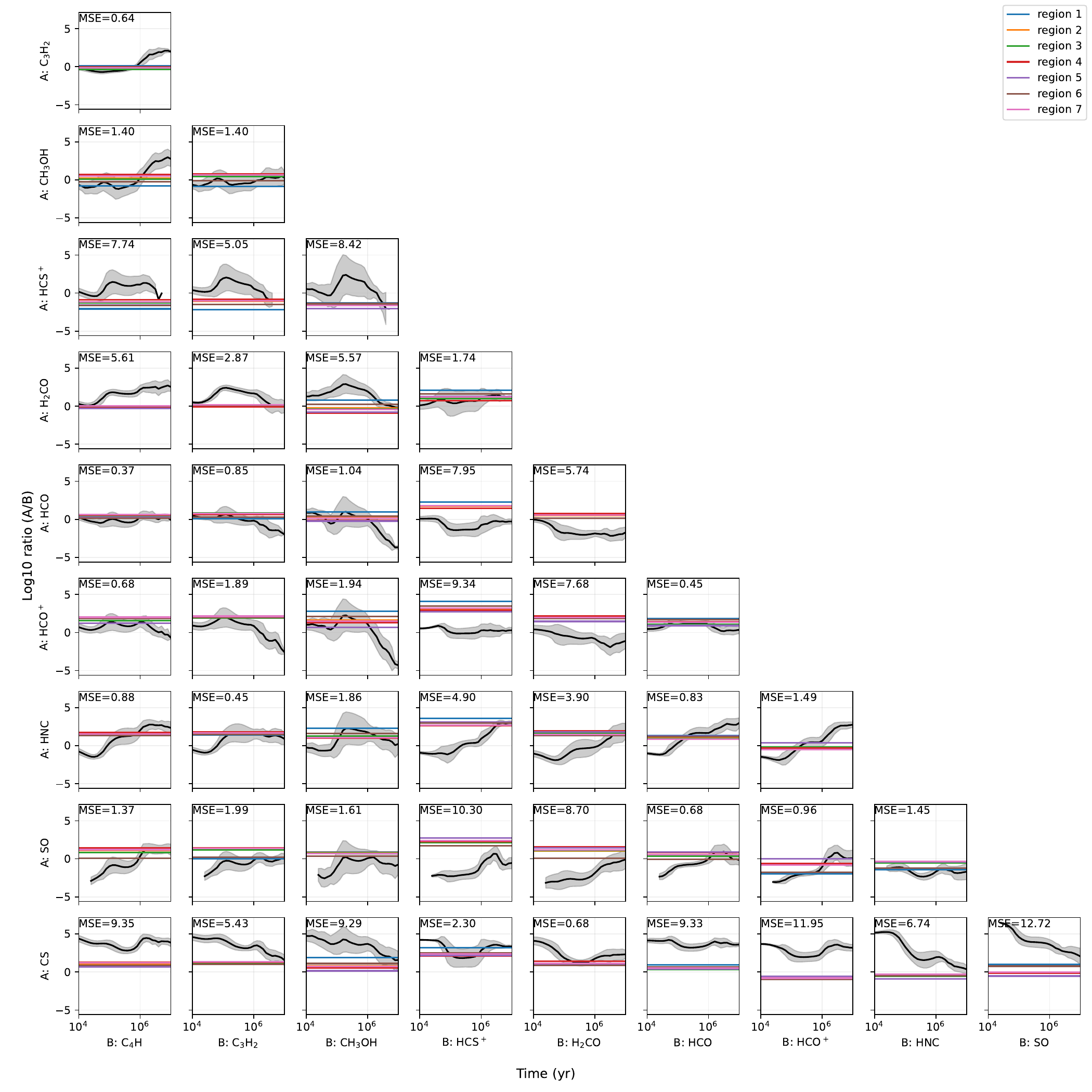}
    \caption{The evolution of the observable ratios as a function of time for the best 50 models, with the
    mean squared error at $10^6$ years.} 
    \label{fig:ratio_evolution}
\end{figure*}

Lastly we investigate the time-dependence of these ratios (Fig.~\ref{fig:ratio_evolution}) in order to gauge whether 
the observed star-forming regions exhibit a younger or older chemical history. 
Figure~\ref{fig:ratio_evolution} shows that for most molecules the fits would not improve at another time, apart from  the 
H$_2$O/\METH, CS/HNC and CS/SO ratios who might benefit from sampling at a later time. 
However this would make the fits for other ratios worse. 
For some molecules there seems to be no time at which the ratios are fit correctly. 
The worst performing molecule, CS, clearly shows an over-prediction in the model as compared to the observations.
However, such inconsistency can be due to a neglected high optical depth in the derivation of the observed column densities. 

Overall, our results indicate that static models as those investigated here can give a range of physical parameters that best match with the observations, but are not optimal to reproduce the set of observed abundances.
A more proper fit for each region would need dynamical modelling, accounting for collapse during the time dependent evolution.
The range of values predicted from the static models suggest that low energetic conditions, in particular low temperatures (20--45~K) and $F_{\rm UV}$ lower than 5 Habing, are favoured, consistent with the location of the source in the Galactic anti-Centre (certainly more "quiescent" than the local and inner Galaxy).
The models also predict that the oxygen elemental abundance should not be smaller than 1/5th of the Solar value, consistent with an extrapolation of the \citet{mendez22} trend which predicts a decrease of a factor 4.6.
The carbon elemental abundance should also not be smaller than $\sim$ 1/5th of the Solar value, but this is well above the value extrapolated from the \citet{mendez22} trend at 23.4~kpc, that is $\sim 1/14$th of the Solar value.
If confirmed by models that include dynamical collapse, such difference would indicate a [C/H] gradient that flattens in the far-outer Galaxy with respect to the trend derived at inner $R_{\rm GC}$. 
Indeed, elemental Galactocentric gradients derived from observations of HII regions in spiral galaxies characterised by extended H I envelopes indicate a flattening of the radial distribution of metals at large galactocentric distances \citep{bresolin17}.
An elemental gradient of carbon flatter than expected in the OG would also explain why the abundance trends with $R_{\rm GC}$ of organics tend to be less steep than the extrapolated gradients \citep{bernal21,fontani22b}.

Another caveat arises from the assumed gas-to-dust mass ratio, which could be higher than that assumed (100) according to the trend proposed by \citet{giannetti17} ($\sim 3000^{+700}_{-2200}$).
We thus performed two tests, one for a model at low density ($\sim 10^3$~\cmc) and one at high density ($\sim 10^6$~\cmc).
In the low density case, the results do not show any significant differences.
In the high density case, some differences are found, but this case does not represent WB670.
Moreover, no firm measurements of this ratio are so far obtained, to our knowledge, at such large $R_{\rm GC}$, and hence assuming a gas-to-dust ratio significantly different from the standard one will introduce a further uncertainty in the models.

\section{Conclusions}
\label{conc}

We used ALMA to observe WB670, the source with the largest Galactocentric distance (23.4~kpc) in CHEMOUT, at a resolution of $\sim 15000$~au. 
We detected emission of \cyclic, C$_4$H, \METH, \FORM, HCO, \HCOpI, HCS$^+$, CS, HN$^{13}$C, and SO, derived their column densities, and compared the observational results with chemical model predictions.
The main results of our study are the following:
\begin{itemize}
\item The molecular emission arises from a filamentary structure oriented SE-NW, where multiple cores are detected.
The filament seems more extended than the ALMA primary beam.
The morphology is different in each tracer. 
The most intense emission of molecular ions, carbon-chain molecules, and \FORM\ is associated with two millimeter continuum, infrared-bright cores.
On the contrary, the \METH\ and SO most intense emission arises predominantly from the part of the filament with no continuum sources.
The narrow linewidths ($\sim 1-2$~\kms) across the filament indicate quiescent gas, despite the presence of the infrared-bright sources;
\item From a LTE analysis of the \METH\ lines, their excitation temperatures are quite low (7--15~K) and could be under-thermally excited.
Derived molecular column densities are comparable with those in local star-forming regions.
There seems to be a spatial anti-correlation between the column density of hydrocarbons, molecular ions, HCO, and \FORM\ on one side, and \METH\ and SO on the other.
This would possibly suggest different formation processes for the two groups of molecules (gas phase processes versus surface processes);
\item \METH\ fractional abundances calculated towards the millimeter continuum cores ($0.4-11\times 10^{-9}$) are consistent with those of the so-called organic-poor cores found in the LMC, where such low \METH\ abundances could be due to an inefficient hydrogenation of CO on grain mantles;
\item Static models that best match the observed column densities favour diffuse gas and low irradiation conditions (expected at large Galactocentric radii), but carbon elemental abundances 3 times higher than that derived extrapolating the [C/H] elemental Galactocentric gradient at 23~kpc. This would indicate a flatter [C/H] trend at large Galactocentric radii, in line with a flat abundance of organics.
Models including dynamical evolution should be able to more properly reproduce the chemical composition of WB670.
\end{itemize}
The results of this work indicate that a proper comparison between observations and models starting from a huge grid of parameters is essential to properly model the chemistry.
Our study would greatly benefit from new observations of more molecular species and more lines at the same (at least) spatial resolution as that obtained here. 
In particular, tracers of the cosmic-ray ionisation rate, basically unconstrained by our study, would be particularly relevant.

\begin{acknowledgements}

This paper makes use of the following ALMA data: ADS/JAO.ALMA$\#$2022.1.00911.S. ALMA is a partnership of ESO (representing its member states), NSF (USA) and NINS (Japan), 
together with NRC (Canada), MOST and ASIAA (Taiwan), and KASI (Republic of Korea), in cooperation with the Republic of Chile.
The Joint ALMA Observatory is operated by ESO, AUI/NRAO and NAOJ. 
F.F., S.V., G.V., and D.G. ackowledge support from the European Research Council (ERC) Advanced grant MOPPEX 833460.
L.C. and V.M.R. acknowledges support from the grant PID2022-136814NB-I00 by the Spanish Ministry of Science, Innovation and Universities/State Agency of Research MICIU/AEI/10.13039/501100011033 and by ERDF, UE; V.M.R. also acknowledge support 
from the grant RYC2020-029387-I funded by MICIU/AEI/10.13039/501100011033 and by "ESF, Investing in your future", and from the Consejo Superior de Investigaciones Cient{\'i}ficas (CSIC) and the Centro de Astrobiolog{\'i}a (CAB) through the project 20225AT015 (Proyectos intramurales especiales del CSIC);
and from the grant CNS2023-144464 funded by MICIU/AEI/10.13039/501100011033 and by “European Union NextGenerationEU/PRTR”.
A.S.-M.\ acknowledges support from the RyC2021-032892-I grant funded by MCIN/AEI/10.13039/501100011033 and by the European Union `Next GenerationEU’/PRTR, as well as the program Unidad de Excelencia María de Maeztu CEX2020-001058-M, and support from the PID2020-117710GB-I00 (MCI-AEI-FEDER, UE).

\end{acknowledgements}

{}

\onecolumn

\begin{appendix}

\section{Maps}
\label{app:maps}

\begin{figure*}[h!]
   \centering
   \includegraphics[width=15cm]{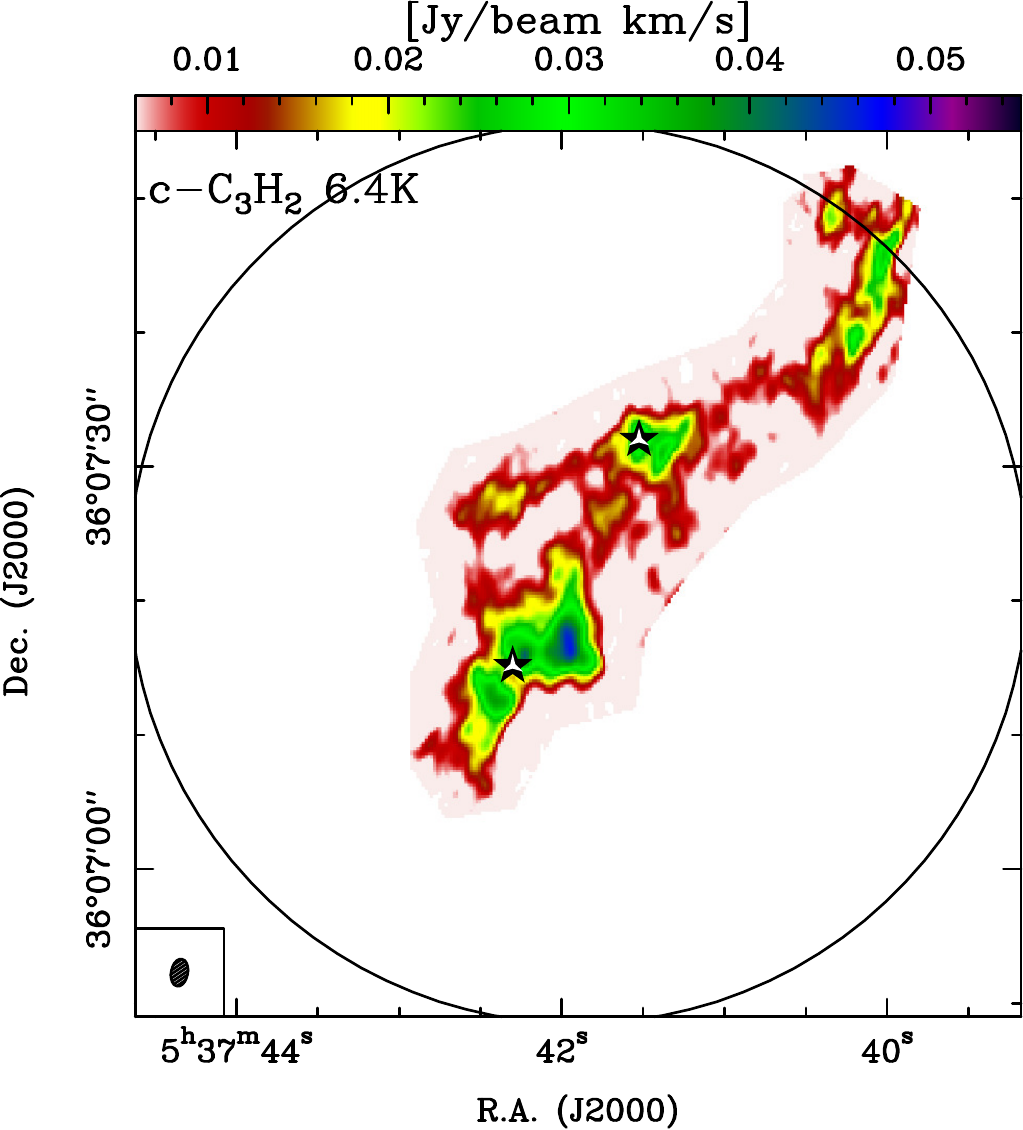}
      \caption{Velocity-integrated emission map of \cyclic\ $J(K_a,K_b)=2(1,2)-1(0,1)$.
      The map is the same shown in Fig.~\ref{fig:morphology}, so we refer to that figure caption for details.}
         \label{fig:mean-1}
   \end{figure*}

\begin{figure*}[h!]
   \centering
   \includegraphics[width=15cm]{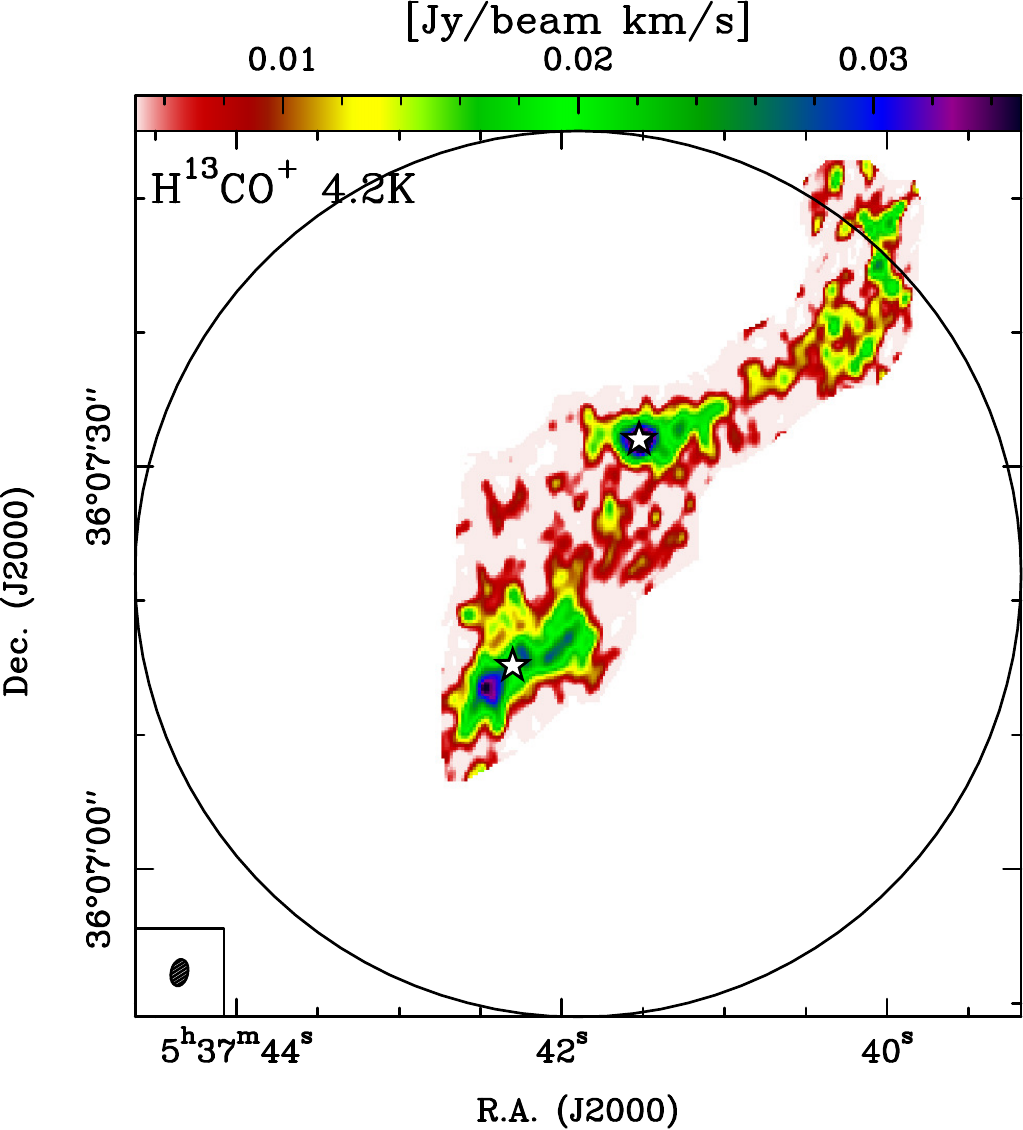}
      \caption{Same as Fig.~\ref{fig:mean-1} for \HCOpI\ $J=1-0$.}
         \label{fig:mean-2}
   \end{figure*}

\begin{figure*}[h!]
   \centering
   \includegraphics[width=15cm]{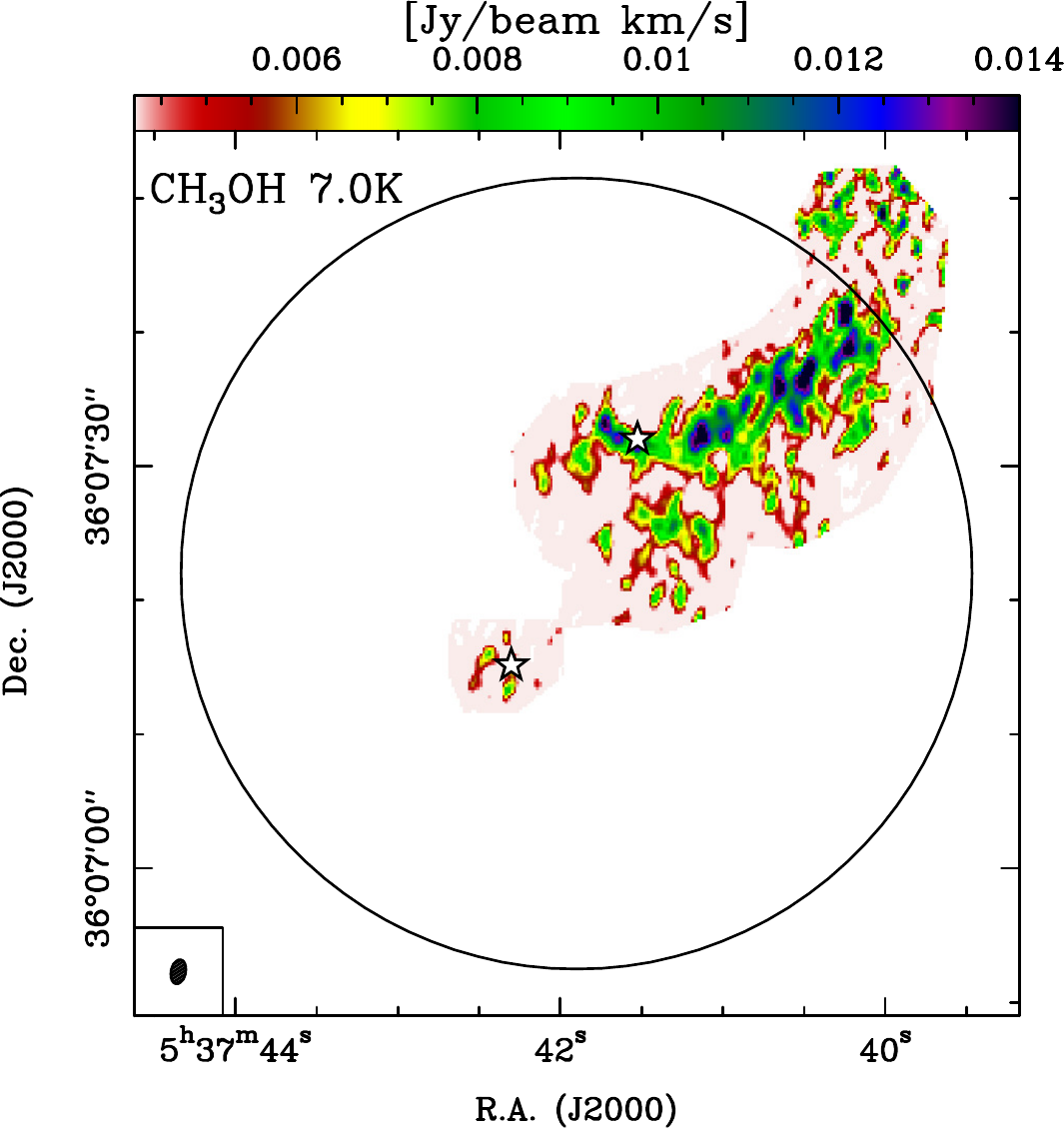}
      \caption{Same as Fig.~\ref{fig:mean-1} for \METH\ $J(K_a,K_b)=2(0,1)-1(0,1) {\rm A}^+$.}
         \label{fig:mean-3}
   \end{figure*}

\begin{figure*}[h!]
   \centering
   \includegraphics[width=15cm]{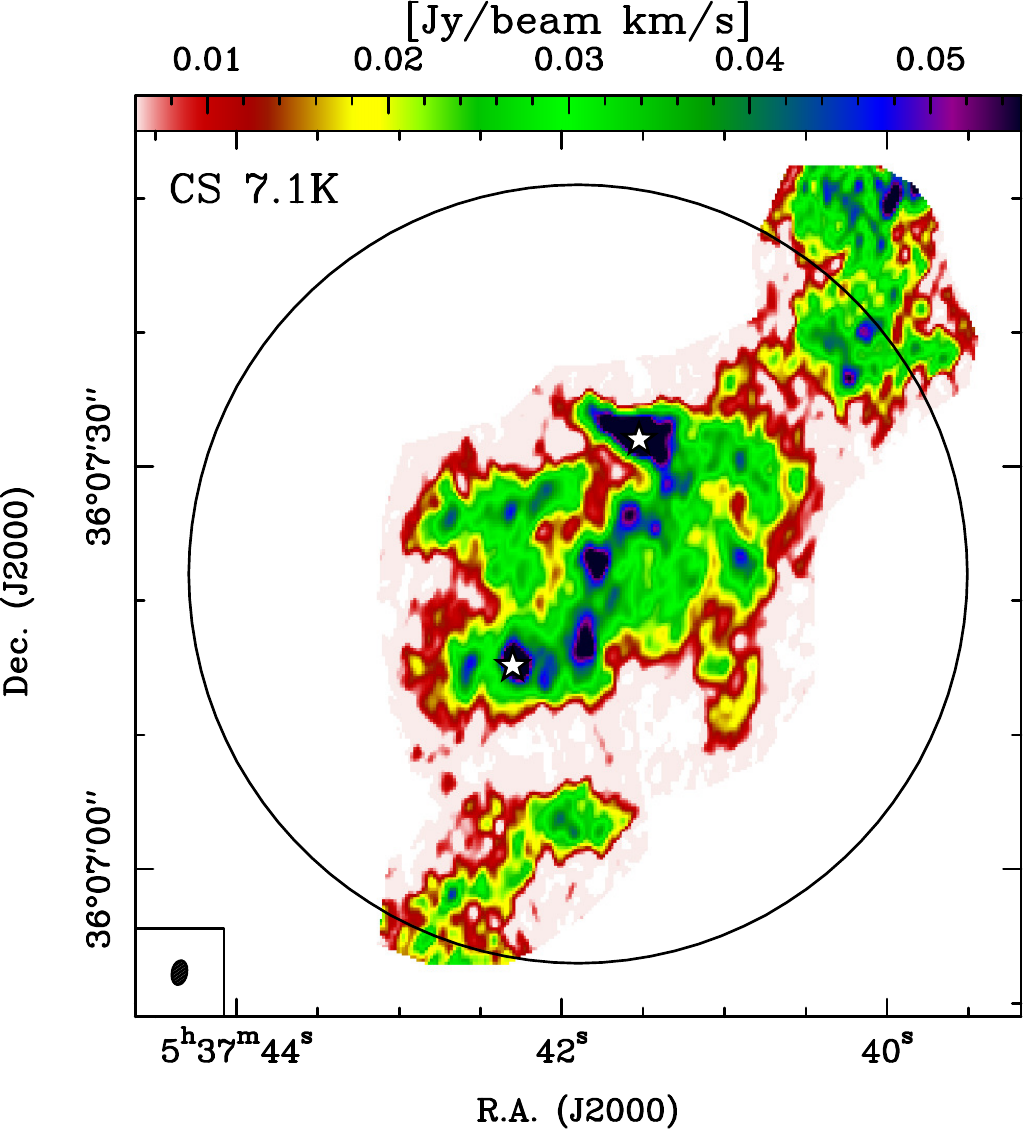}
      \caption{Same as Fig.~\ref{fig:mean-1} for CS $J=2-1$.}
         \label{fig:mean-4}
   \end{figure*}

\begin{figure*}[h!]
   \centering
   \includegraphics[width=15cm]{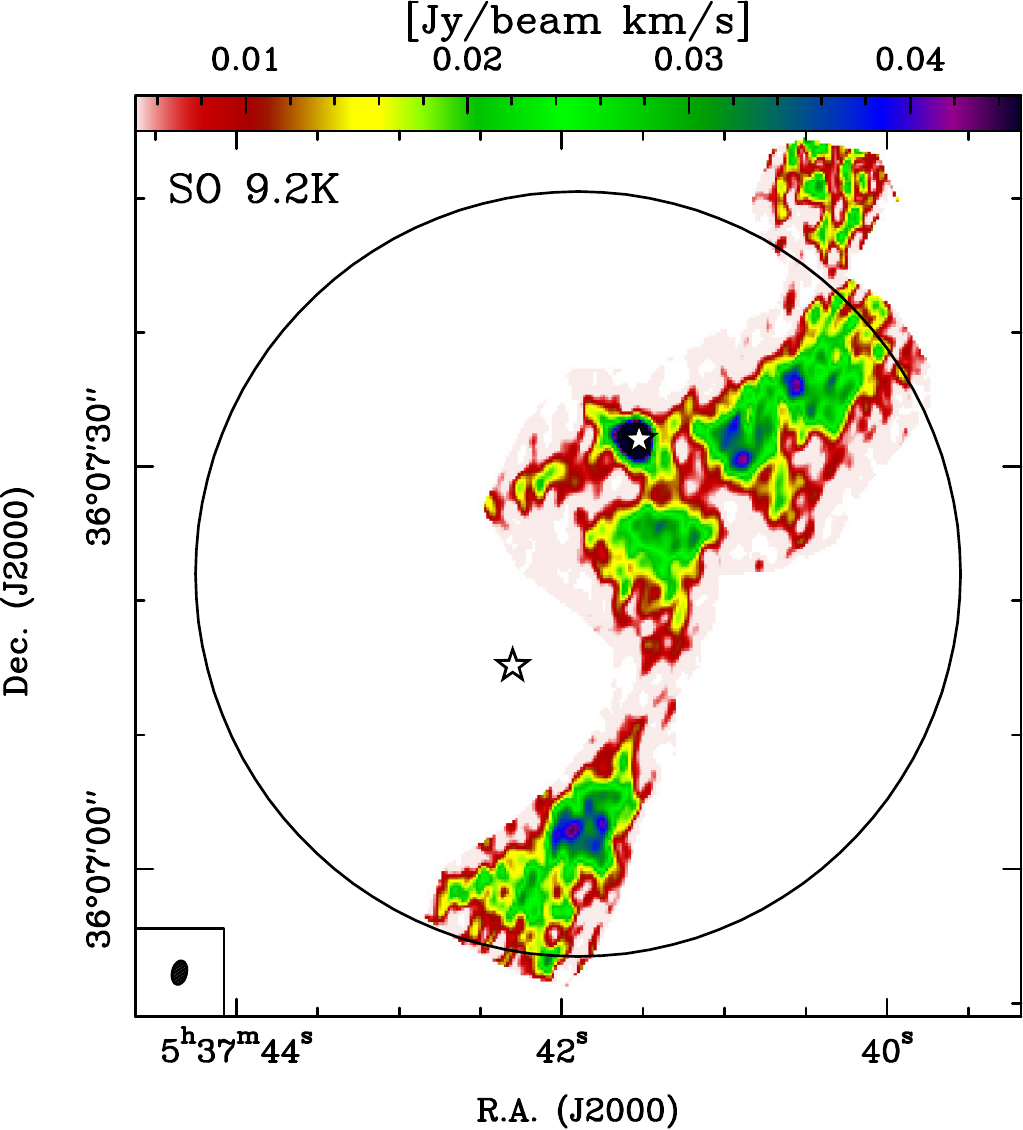}
      \caption{Same as Fig.~\ref{fig:mean-1} for SO $J(K)=3(2)-2(1)$.}
         \label{fig:mean-5}
   \end{figure*}

\begin{figure*}[h!]
   \centering
   \includegraphics[width=15cm]{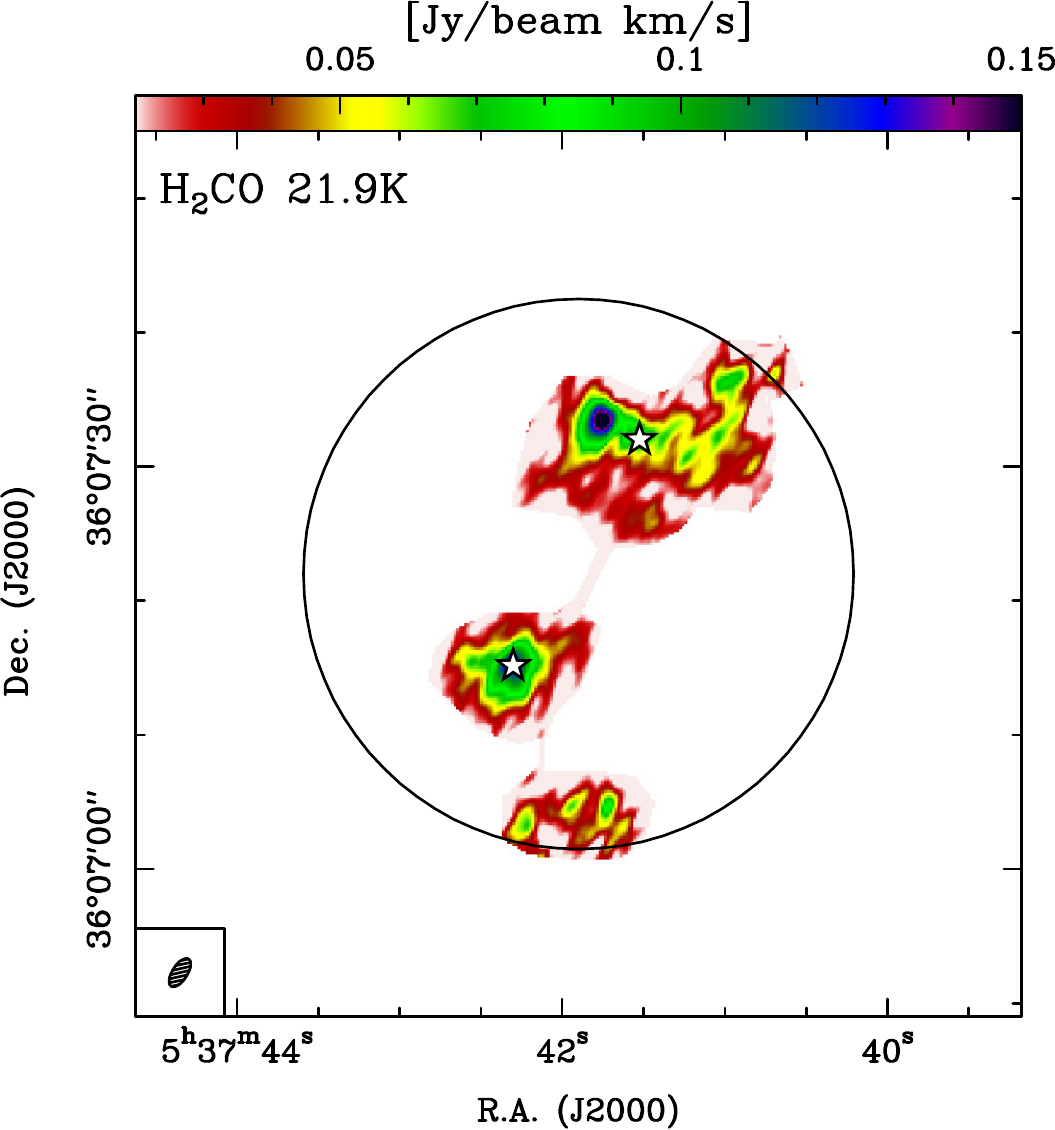}
      \caption{Same as Fig.~\ref{fig:mean-1} for \FORM\ $J(K_a,K_b)=2(1,2)-1(1,1)$.}
         \label{fig:mean-6}
   \end{figure*}

The maps of the integrated emission in Fig.~\ref{fig:morphology} are shown, enlarged, in Figures~\ref{fig:mean-1}--\ref{fig:mean-6}.

\FloatBarrier

   \FloatBarrier

   \FloatBarrier

   \FloatBarrier

  \FloatBarrier

     \FloatBarrier

\FloatBarrier
\section{Spectra}
\label{app:spectra}

Spectra of the detected lines, extracted from the regions identified in Sect.~\ref{maps-mol}.

\FloatBarrier
   \begin{figure*}[h!]
   \centering
   \includegraphics[width=13cm]{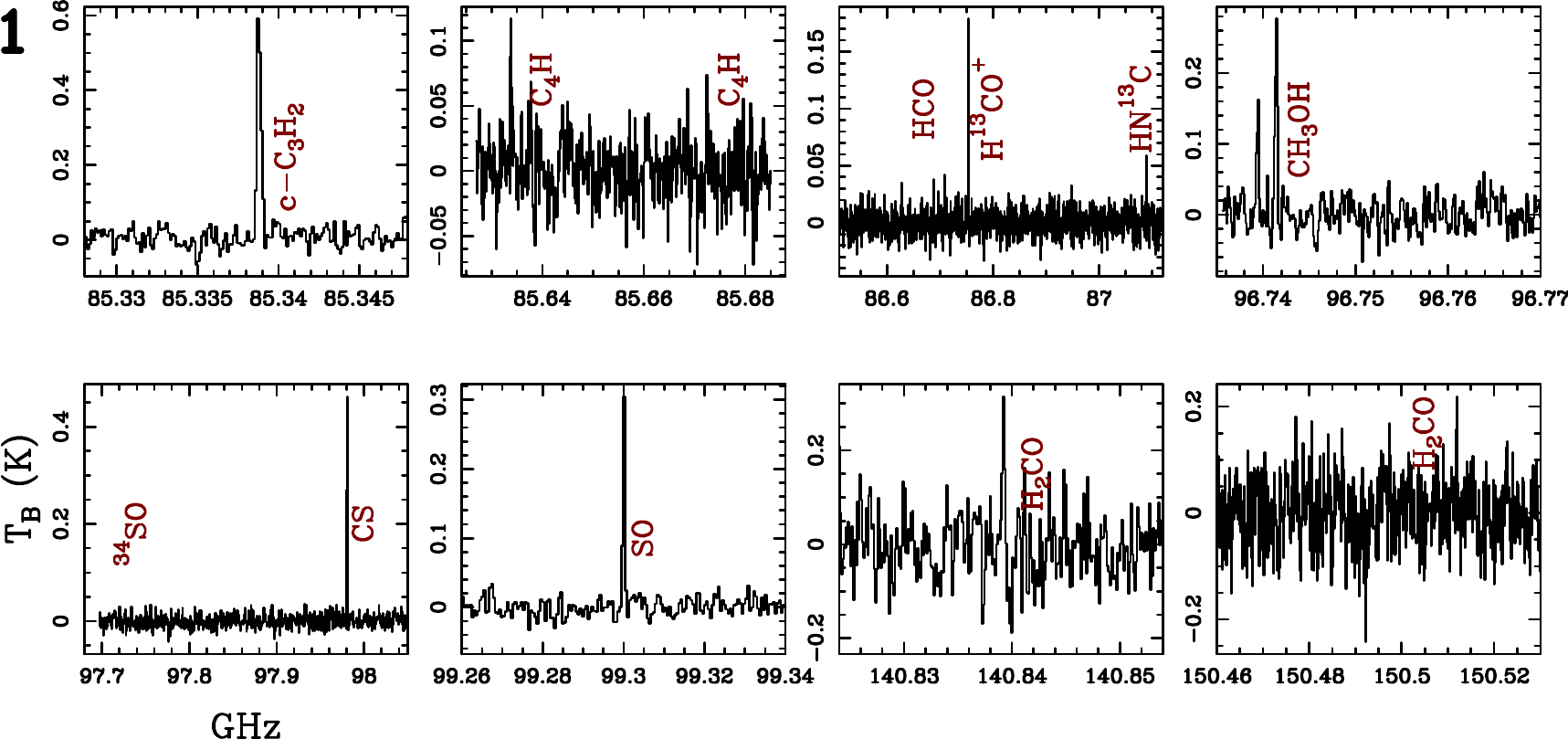}
   \includegraphics[width=13cm]{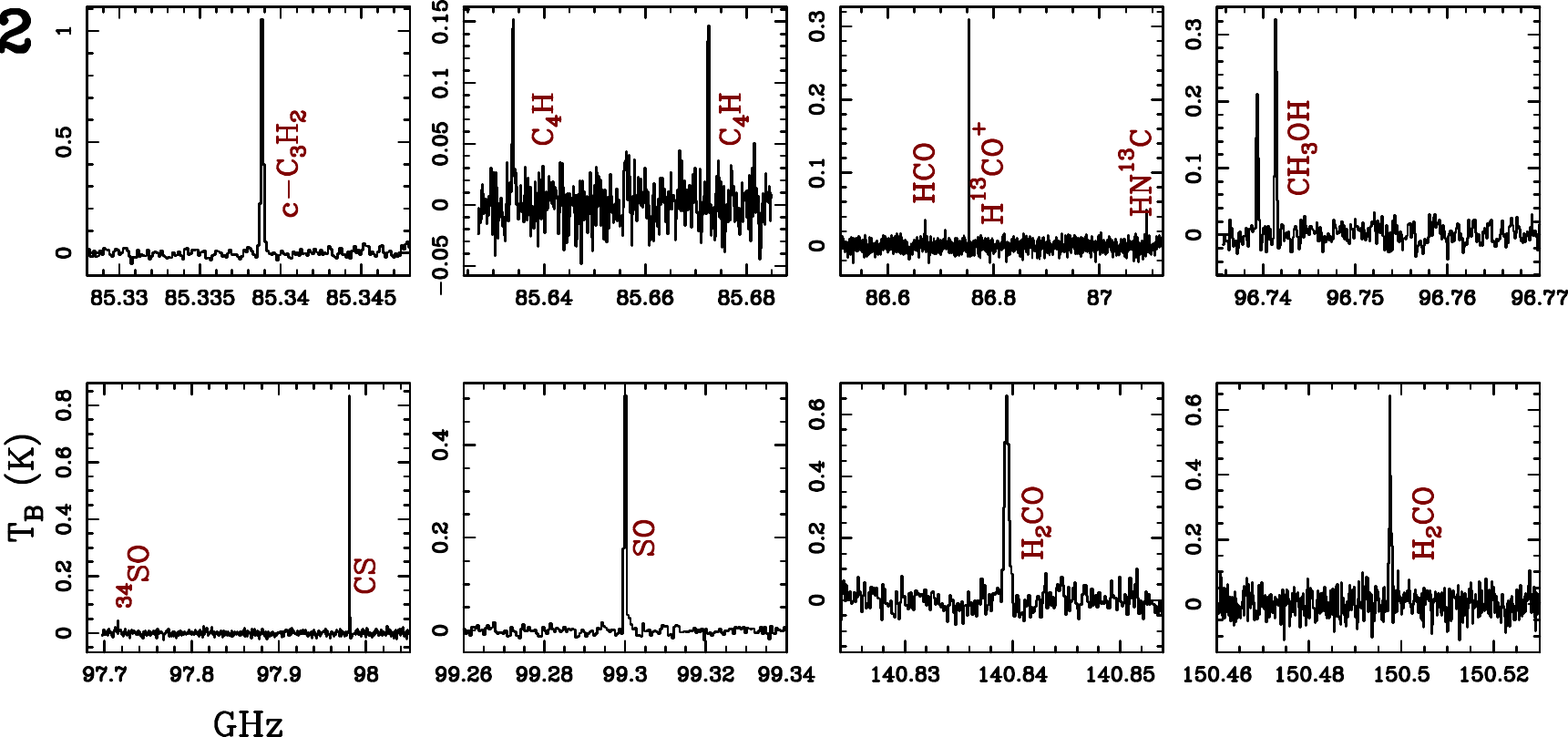}
   \includegraphics[width=13cm]{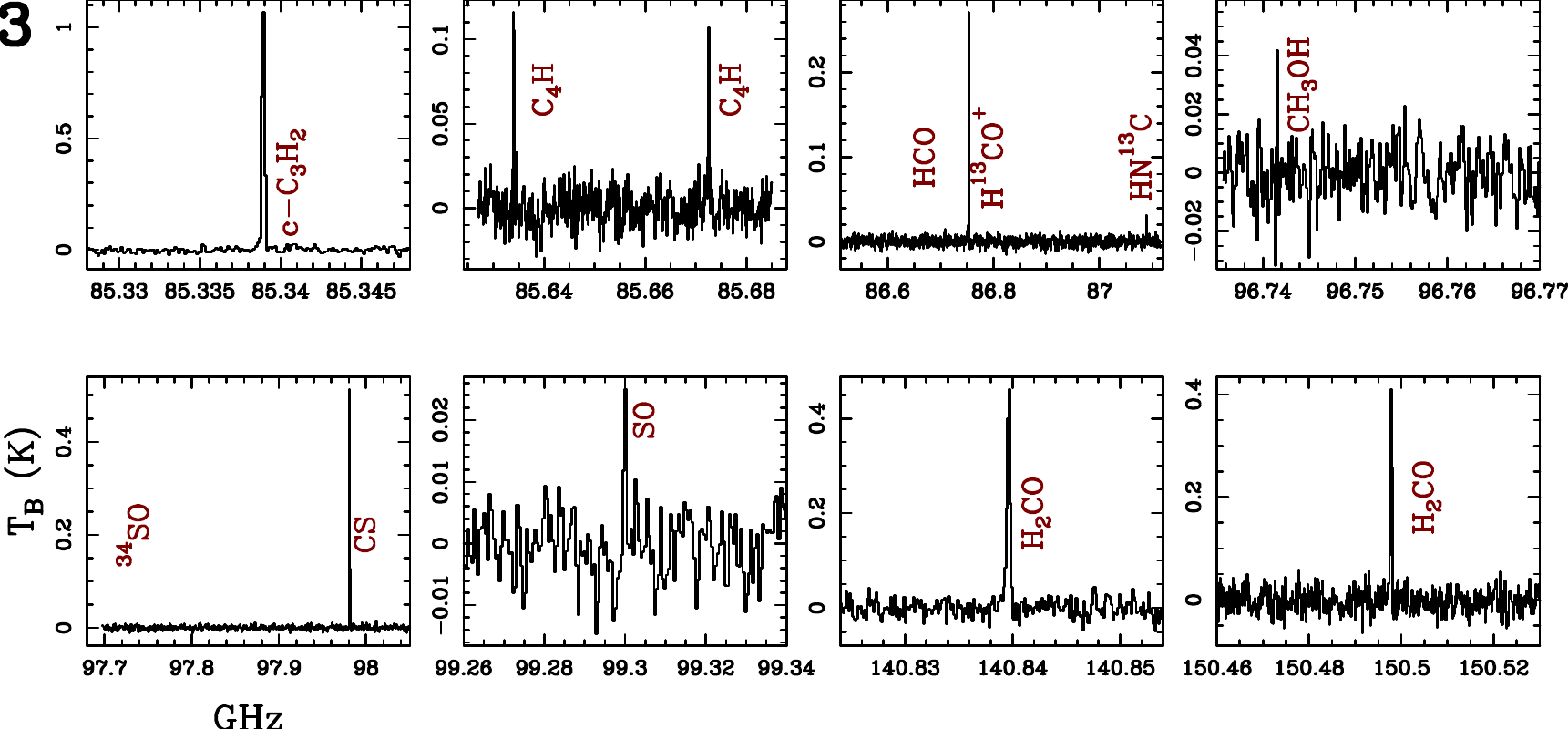}
      \caption{Spectra extracted from the \cyclic\ region 1, 2, and 3, from top to bottom, in brightness temperature (\Tb) units. 
      On the x-axis we show the rest frequency.}
         \label{fig:spec-C3H2}
   \end{figure*}

\FloatBarrier
   \begin{figure*}[h!]
   \centering
   \includegraphics[width=13cm]{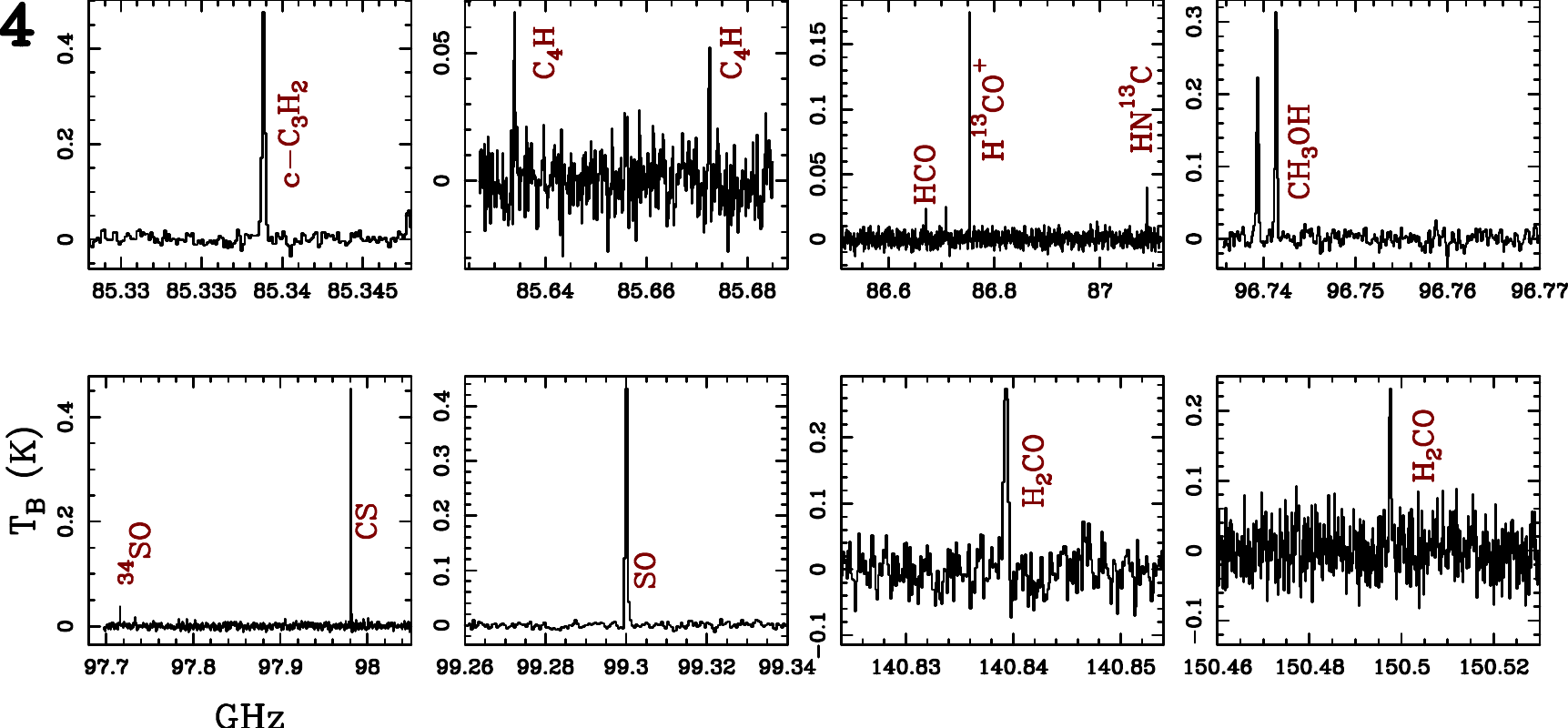}
   \includegraphics[width=13cm]{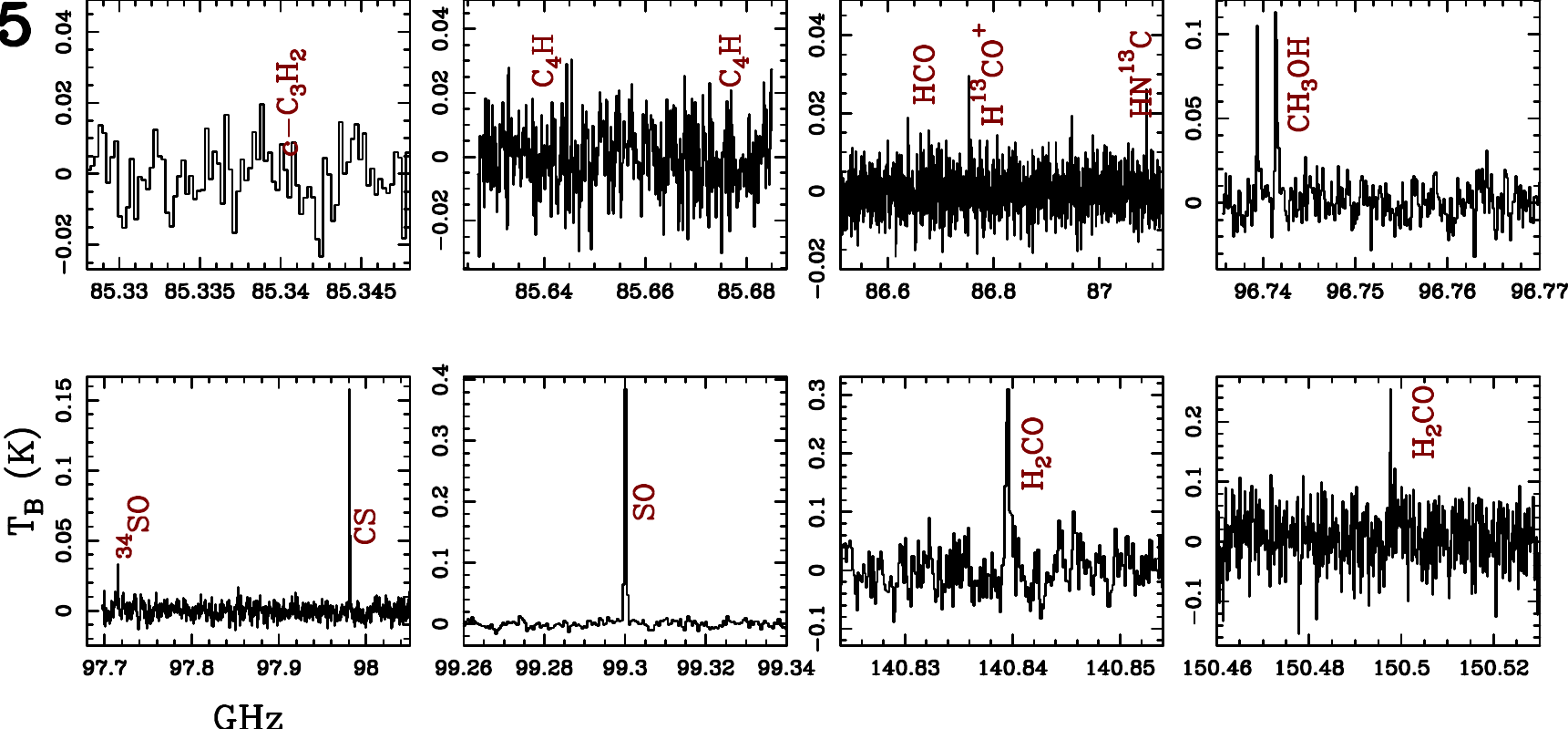}
   \includegraphics[width=13cm]{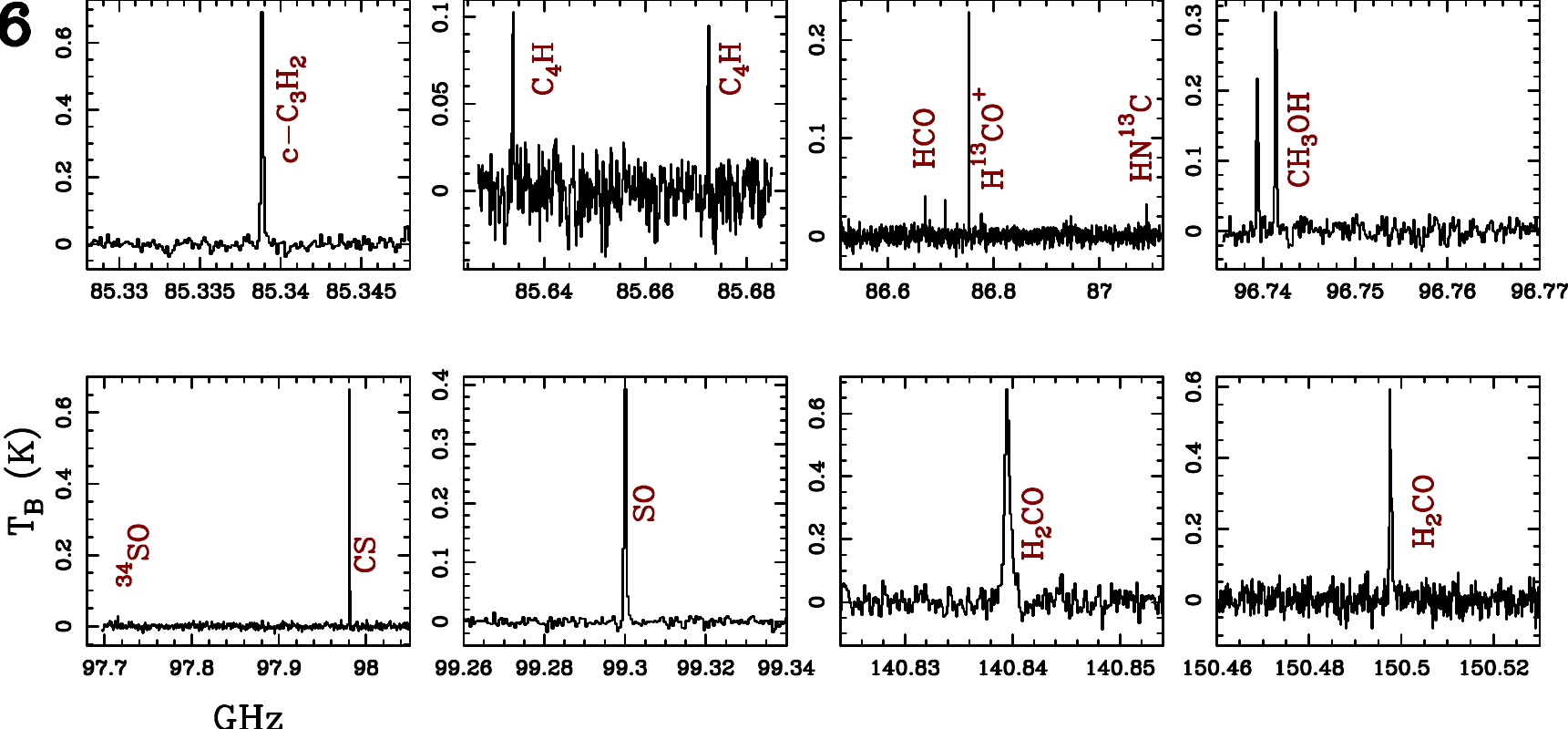}
   \includegraphics[width=13cm]{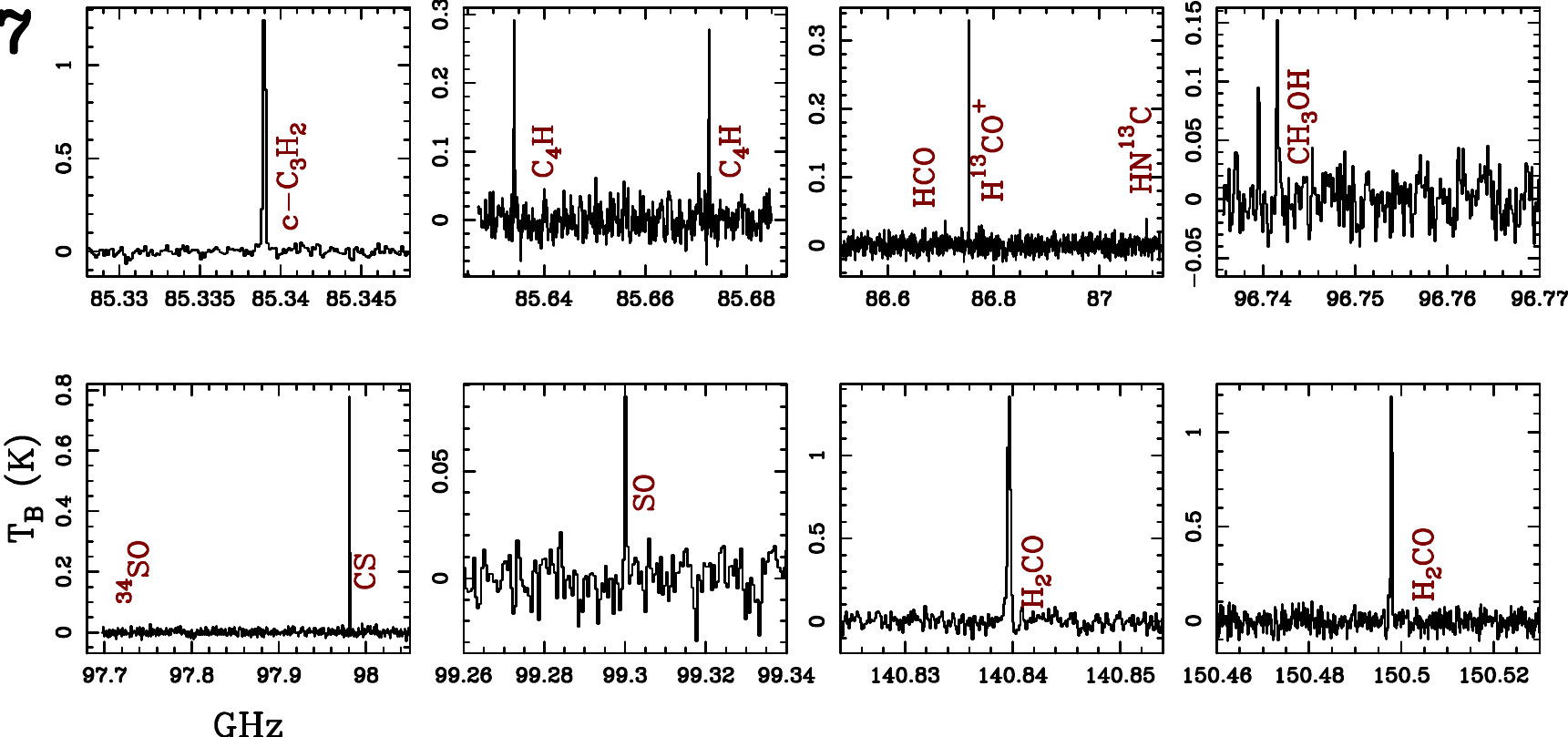}   
      \caption{Same as Fig.~\ref{fig:spec-C3H2} for the regions 4, 5, 6, and 7, from top to bottom.}
         \label{fig:spec-H2CO}
   \end{figure*}

\newpage

\FloatBarrier
\section{Fit results}
\label{app:fits}

Fit results obtained with {\sc madcuba} and {\sc class} towards the lines detected in the seven molecular emitting regions identified in Fig.~\ref{fig:morphology}.

\FloatBarrier
   \begin{table}[h!]
   \begin{center}
\caption{\label{tab:fits-1} Fit results of the molecular lines identified in region 1.}
\begin{tabular}{l c c c c }
\hline
\hline
molecule  &  \Ntot\  & \Tex\ & $v_{\rm LSR}$  & FWHM \\
          &  ($10^{12}$\cmq)  & (K)   & (\kms )        & (\kms) \\
\hline
\multicolumn{5}{c}{{\sc madcuba}\tablefootmark{a}} \\
\hline
C$_4$H     & 5.1(0.3)  & 6.4   &  -17.02(0.09) &  0.79(0.05) \\
\cyclic\   & 6.5(0.2)  & 6.4   &  -17.70(0.01) &  0.75(0.02) \\
\METH\     & 0.9(0.3)  & 6.4(0.9)  & -18.1(0.2)    &  0.7   \\
HCS$^+$    & 0.04(0.01)   & 6.4   &  -17.1(0.08)  & 0.7   \\
\FORM\     & 5(1)     & 6.4   &  -17.1     &  0.7   \\
\hline
\multicolumn{5}{c}{{\sc class}\tablefootmark{c}} \\
\hline
molecule  &  \Ntot\  & $\int T_{\rm B}{\rm d}v$ & $v_{\rm LSR}$  & FWHM \\
          &  ($10^{12}$\cmq)  & (K)   & (\kms )        & (\kms) \\
\hline
HCO\tablefootmark{c}  & $\leq 1.9$    & --   &  --   & --  \\ 
\HCOpI\    & 8.7(1.1) & 0.59(0.02)  &  --17.82(0.05)   & 1.89(0.09) \\ 
HN$^{13}$C & 2.8(0.7) & 0.09(0.01)  &  --17.4(0.2)  & 2.3(0.6) \\ 
SO         & 6.3(1.9) & 0.06(0.01)  &  --17.9(0.1)  & --\tablefootmark{d} \\  
CS         & 68(7) & 1.18(0.01)  &  --17.28(0.01)  & 2.16(0.02)  \\  
\hline
   \end{tabular}
   \end{center}
\tablefoot{Line parameters without uncertainties (quoted in brackets) were fixed in the fit. The error on \Ntot\ contains the calibration error of $\sim 10\%$ when derived with {\sc class}, while does not when derived with {\sc madcuba}. Hence, in this case the total uncertainty is given by the quadrature sum of the quoted error and of a $10\%$ calibration error.
\tablefoottext{a}{Parameters obtained fitting the lines with {\sc madcuba};}
\tablefoottext{b}{Parameters obtained fitting the lines with {\sc class};}
\tablefoottext{c}{Parameters obtained fitting with a Gaussian profile the hyperfine component $F=1-0$, detected in almost all regions except regions 1 and 5;}
\tablefoottext{d}{Parameter not constrained because the uncertainty provided by the fit is higher than the output value.}
}
   \end{table}

\FloatBarrier
   \begin{table}[h!]
   \begin{center}
\caption{\label{tab:fits-2} Same as Table~\ref{tab:fits-1} for region 2.}
\begin{tabular}{l c c c c }
\hline
\hline
\multicolumn{5}{c}{{\sc madcuba}\tablefootmark{a}} \\
\hline
molecule  &  \Ntot\  & \Tex\ & $v_{\rm LSR}$  & FWHM \\
          &  ($10^{12}$\cmq)  & (K)   & (\kms )        & (\kms) \\
\hline
C$_4$H     & 7.6(0.6)   &    8.3   & --17.14(0.02)  & 0.76(0.08) \\ 
\cyclic\   & 5.1(0.2)   &    8.3   & --17.39(0.01)  & 0.62(0.02) \\
\METH\     & 13.5(3)    &    8.3(1.3) &  --17.56(0.03) &  0.74(0.06) \\
HCS$^+$    & 0.5(0.1)   &    8.3   & --17.4(0.2)  & 1.70(0.5)  \\
\FORM\     & 8(1)    &    8.3   & --17.5      & 1.0     \\
\hline
\multicolumn{5}{c}{{\sc class}\tablefootmark{b}} \\
\hline
molecule  &  \Ntot\  & $\int T_{\rm B}{\rm d}v$ & $v_{\rm LSR}$  & FWHM \\
          &  ($10^{12}$\cmq)  & (K)   & (\kms )        & (\kms) \\
\hline
HCO\tablefootmark{c}     & 6(5)   & 0.15(0.1) & --17(2) & --\tablefootmark{d} \\
\HCOpI\    & 8.6(1.2)   & 0.62(0.03) & --17.38(0.06) & 1.86(0.08) \\
HN$^{13}$C & 3.7(1.0)   & 0.13(0.02) & --17.5(0.1) & --\tablefootmark{d} \\
SO         & 91(10)   & 1.06(0.02) & --18.07(0.03) & 1.79(0.05) \\
CS         & 84(9)   & 1.75(0.02) & --17.00(0.04) & 1.81(0.03) \\
\hline
   \end{tabular}
   \end{center}
   \end{table}

\FloatBarrier
   \begin{table}[h]
   \begin{center}
\caption{\label{tab:fits-3} Same as Table~\ref{tab:fits-1} for region 3.}
\begin{tabular}{l c c c c }
\hline
\hline
\multicolumn{5}{c}{{\sc madcuba}\tablefootmark{a}} \\
\hline
molecule  &  \Ntot\  & \Tex\ & $v_{\rm LSR}$  & FWHM \\
          &  ($10^{12}$\cmq)  & (K)   & (\kms )        & (\kms) \\
\hline
C$_4$H     & 10(2) &  6.4       &  --17.02(0.09) &  1.15(0.2) \\
\cyclic\   & 4.6(0.4)  & 6.4    & --17.18(0.04) & 1.1(0.1)\\
\METH\     & 12(5)  & 6.4(1.3) &  --17.71(0.04) &  0.95(0.09) \\
HCS$^+$    & 0.5(0.08)   & 6.4  & --17.45(0.08) & 1.1(0.2)\\
\FORM\     & 5.0(1.0)     & 6.4 &  --17.5  & 1.0   \\
\hline
\multicolumn{5}{c}{{\sc class}\tablefootmark{b}} \\
\hline
molecule  &  \Ntot\  & $\int T_{\rm B}{\rm d}v$ & $v_{\rm LSR}$  & FWHM \\
          &  ($10^{12}$\cmq)  & (K)   & (\kms )        & (\kms) \\
\hline
HCO\tablefootmark{c}   & 8(4)   & 0.17(0.06) & --17.0(0.7) & 3.8(1.6) \\
\HCOpI\    & 5.7(1.0)  & 0.39(0.03) & --17.2(0.1) & 1.8(0.2) \\
HN$^{13}$C & 3.8(1.4)  & 0.12(0.03) & --17.9(0.5) & 2.0(0.5) \\
SO         & 69(10)  & 0.61(0.03) & --18.14(0.08) & 1.8(0.1) \\
CS         & 65(8)  & 1.13(0.03) & --16.79(0.04) & 1.8(0.2) \\
\hline
   \end{tabular}
   \end{center}
   \end{table}

\FloatBarrier
   \begin{table}[h]
   \begin{center}
\caption{\label{tab:fits-4} Same as Table~\ref{tab:fits-1} for region 4.}
\begin{tabular}{l c c c c }
\hline
\hline
\multicolumn{5}{c}{{\sc madcuba}\tablefootmark{a}} \\
\hline
molecule  &  \Ntot\  & \Tex\ & $v_{\rm LSR}$  & FWHM \\
          &  ($10^{12}$\cmq)  & (K)   & (\kms )        & (\kms) \\
\hline
C$_4$H     & 2.9(0.4)  & 8.4      & --17.25(0.05) & 0.86(0.13) \\
\cyclic\   & 2.6(0.1)  & 8.4      & --17.38(0.01) & 0.76(0.03) \\
\METH\     & 16(3)  & 8.4(1.2) & --17.59(0.03) & 0.83(0.06) \\
HCS$^+$    & 0.4(0.05)  & 8.4      & --17.2  & 0.8  \\
\FORM\     & 2.0(0.3)  & 8.4      & --17.1  & 0.7  \\
\hline
\multicolumn{5}{c}{{\sc class}\tablefootmark{b}} \\
\hline
molecule  &  \Ntot\  & $\int T_{\rm B}{\rm d}v$ & $v_{\rm LSR}$  & FWHM \\
          &  ($10^{12}$\cmq)  & (K)   & (\kms )        & (\kms) \\
\hline
HCO\tablefootmark{c}   & 2.7(1.8) &  0.07(0.004)  & --16.2(0.8)  & 2.5(1.3) \\
\HCOpI\    & 5.0(0.7) & 0.36(0.01)  & --17.54(0.05)  & 1.92(0.06) \\          
HN$^{13}$C & 2.6(0.6) &  0.09(0.01) & --17.4(0.1) & 1.7(1.3) \\
SO         & 75(9) &  0.88(0.01) & --18.17(0.02) & 1.83(0.03) \\
CS         & 50(6) &  1.06(0.01) & --16.86(0.02) & 1.85 (0.05) \\
\hline
   \end{tabular}
   \end{center}
   \end{table}

\FloatBarrier
   \begin{table}[h]
   \begin{center}
\caption{\label{tab:fits-5} Same as Table~\ref{tab:fits-1} for region 5.}
\begin{tabular}{l c c c c }
\hline
\hline
\multicolumn{5}{c}{{\sc madcuba}\tablefootmark{a}} \\
\hline
molecule  &  \Ntot\  & \Tex\ & $v_{\rm LSR}$  & FWHM \\
          &  ($10^{12}$\cmq)  & (K)   & (\kms )        & (\kms) \\
\hline
C$_4$H     & $\leq 3.2$   & 15     & --    & -- \\
\cyclic\   & --\tablefootmark{e}   & --    & --    & -- \\
HCS$^+$    & $\leq 0.09$  & 15     & --   & --      \\
\FORM\     & 1.7(0.3)   & 15     & --17.64(0.07)   & 0.9(0.2)      \\
\hline
\multicolumn{5}{c}{{\sc class}\tablefootmark{b}} \\
\hline
molecule  &  \Ntot\  & $\int T_{\rm B}{\rm d}v$ & $v_{\rm LSR}$  & FWHM \\
          &  ($10^{12}$\cmq)  & (K)   & (\kms )        & (\kms) \\
\hline
\METH\     & 10(1)         & 15    & --17.66(0.04)\tablefootmark{f} & 0.9(0.1)\tablefootmark{f} \\
HCO\tablefootmark{c}    & $\leq 1.5$   & --  &  --  & -- \\
\HCOpI\    & 0.8(0.3)  &  0.05(0.01)  &  --17.6(0.5)  &  1.7(1.5)  \\         
HN$^{13}$C & 1.8(0.6)  &  0.06(0.01)  &  --17.4(0.2)  &  --\tablefootmark{d}    \\
SO         & 48(6)  &  0.72(0.01)  &  --18.33(0.03)  &  1.76(0.03)   \\
CS         & 14(2)  &  0.35(0.01)  &  --17.55(0.05)  &  2.0(0.1)  \\
\hline
   \end{tabular}
   \end{center}
\tablefoot{
\tablefoottext{e}{Fit not performed because the line shows self-absorption at line centre;}
\tablefoottext{f}{Derived from the line at 96.7414~GHz.}
}
   \end{table}

\FloatBarrier
   \begin{table}[h]
   \begin{center}
\caption{\label{tab:fits-6} Same as Table~\ref{tab:fits-1} for region 6.}
\begin{tabular}{l c c c c }
\hline
\hline
\multicolumn{5}{c}{{\sc madcuba}\tablefootmark{a}} \\
\hline
molecule  &  \Ntot\  & \Tex\ & $v_{\rm LSR}$  & FWHM \\
          &  ($10^{12}$\cmq)  & (K)   & (\kms )        & (\kms) \\
\hline
C$_4$H     & 10(0.6)   & 9     & --17.72(0.02)    & 0.69(0.05) \\
\cyclic\   & 7.6(0.2)   & 9     & --17.91(0.01)    & 0.68(0.02) \\
\METH\     & 6(2)      & 9.4(2.3) & --18.12(0.04) & 0.62(0.08) \\
HCS$^+$    & 0.23       & 9     & --17.3   & 0.9       \\
\FORM\     & 10.0(1.1)   & 9     & --17.8   & 0.9       \\
\hline
\multicolumn{5}{c}{{\sc class}\tablefootmark{b}} \\
\hline
molecule  &  \Ntot\  & $\int T_{\rm B}{\rm d}v$ & $v_{\rm LSR}$  & FWHM \\
          &  ($10^{12}$\cmq)  & (K)   & (\kms )        & (\kms) \\
\hline
HCO\tablefootmark{c}    & 4(3)  &   0.09(0.06)   & --16.9(0.7)  &  --\tablefootmark{d}  \\
\HCOpI\    & 11.4(1.5)  &  0.82(0.02)  &  --17.07(0.02)  &  1.7(0.6)  \\         
HN$^{13}$C & 3.5(1.1)  &  0.12(0.03)  &  --17.3(0.3)  &  2.7(0.7)    \\
SO         & 12(3)  &  0.16(0.02)  &  --18.5(0.2)  &  1.7(0.2)   \\
CS         & 73(8)  &  1.63(0.02)  &  --17.65(0.02)  &  1.79(0.04)  \\
\hline
   \end{tabular}
   \end{center}
   \end{table}

\FloatBarrier
   \begin{table}[h]
   \begin{center}
\caption{\label{tab:fits-7} Same as Table~\ref{tab:fits-1} for region 7.}
\begin{tabular}{l c c c c }
\hline
\hline
\multicolumn{5}{c}{{\sc madcuba}\tablefootmark{a}} \\
\hline
molecule  &  \Ntot\  & \Tex\ & $v_{\rm LSR}$  & FWHM \\
          &  ($10^{12}$\cmq)  & (K)   & (\kms )        & (\kms) \\
\hline
C$_4$H     & 4.4(0.4) &  8.4   & --17.18(0.03) & 0.73(0.09)   \\
\cyclic\   & 3.2(0.1) &  8.4   & --17.40(0.01) & 0.60(0.02)   \\
\METH\     & 14(3) &  8.4(1.3) & --17.59(0.03) & 0.73(0.06)   \\
HCS$^+$    & 0.3(0.06) &  8.4   & --17.3(0.1)  & 0.9(0.2)    \\
\FORM\     & 5(1) &  8.4   & --17.5     & 0.8       \\
\hline
\multicolumn{5}{c}{{\sc class}\tablefootmark{b}} \\
\hline
molecule  &  \Ntot\  & $\int T_{\rm B}{\rm d}v$ & $v_{\rm LSR}$  & FWHM \\
          &  ($10^{12}$\cmq)  & (K)   & (\kms )        & (\kms) \\
\hline
HCO\tablefootmark{c}   & 5(3)  & 0.12(0.06) & --16.0(0.8) & 2.9(1.9) \\
\HCOpI\    & 7.0(1.0)  & 0.50(0.02) & --17.48(0.06) & 2.07(0.08) \\
HN$^{13}$C & 2.3(0.7)  & 0.08(0.01) & --17.4(0.1) & --\tablefootmark{d} \\
SO         & 70(8)  & 0.82(0.01) & --18.16(0.02) & 1.89(0.04) \\
CS         & 69(8)  & 1.45(0.02) & --17.14(0.01) & 2.00(0.02) \\
\hline
   \end{tabular}
   \end{center}
   \end{table}

\end{appendix}

\end{document}